\documentclass[letterpaper,twocolumn,10pt,dvipsnames]{article}
\usepackage{usenix-2020-09}

\usepackage{url}
\usepackage{comment}
\usepackage{graphicx}
\usepackage{subfig}
\usepackage{balance}
\usepackage{color}
\usepackage{hyperref}
\usepackage{url}

\usepackage{breakurl}
\usepackage{enumitem}

\usepackage{lipsum}
\usepackage{titlesec}

\usepackage{algorithm}
\usepackage{algorithmic}
\usepackage{amsmath}
\usepackage{amssymb}

\usepackage{multirow}

\usepackage{booktabs}

\usepackage{graphicx}

\usepackage{authblk}
\usepackage{color}

\newcommand*{\red}[1]{\textcolor{Red}{#1}}
\newcommand*{\blue}[1]{\textcolor{RoyalBlue}{#1}}
\newcommand*{\green}[1]{\textcolor{Green}{#1}}

 \newcommand{\authnote}[2]{{\bf \textcolor{Green}{#1}: \em \textcolor{Green}{#2}}}
 \newcommand{\yizheng}[1]{\authnote{Yizheng}{#1}}

\begin{document}

\title{Continuous Learning for Android Malware Detection}

\author{Yizheng Chen, Zhoujie Ding, and David Wagner\\ \emph{UC Berkeley}}

%%%%% END AUTHORS

\maketitle

\begin{abstract}

Machine learning methods can detect Android malware with very high accuracy. However, these classifiers have an Achilles heel, concept drift: they rapidly become out of date and ineffective, due to the evolution of malware apps and benign apps. Our research finds that, after training an Android malware classifier on one year’s worth of data, the F1 score quickly dropped from 0.99 to 0.76 after 6 months of deployment on new test samples.

In this paper, we propose new methods to combat the concept drift problem of Android malware classifiers.
Since machine learning technique needs to be continuously deployed, we use active learning: we select new samples for analysts to label, and then add the labeled samples to the training set to retrain the classifier. 
Our key idea is, similarity-based uncertainty is more robust against concept drift.
Therefore, we combine contrastive learning with active learning.
We propose a new hierarchical contrastive learning scheme, and a new sample selection technique
to continuously train the Android malware classifier.
Our evaluation shows that this leads to significant improvements, compared to previously published methods for active learning.
Our approach reduces the false negative rate from 14\% (for the best baseline) to 9\%, while also reducing the false positive rate (from 0.86\% to 0.48\%).
Also, our approach maintains more consistent performance across a seven-year time period than past methods.

\end{abstract}

\section{Introduction}

Machine learning for Android malware detection has been deployed in industry.
However, these classifiers have an Achilles heel, concept drift: they rapidly
become out of date and ineffective. Concept drift happens for many reasons.
For example, malware authors may add new malicious functionality, modify their apps to evade detection,
or create new types of malware that’s never been seen before, and benign apps regularly
release updates.
Our research finds that, after training an Android malware classifier on
one year’s worth of data, the classifier's F1 score quickly dropped from 0.99 to 0.76
after 6 months of deployment on new test samples.

Therefore, rather than learning a single, fixed classifier, security applications require continuous learning, where the classifier is continuously updated to keep up with concept drift.
The state-of-the-art solutions to combat concept drift use active learning to adapt to concept drift.
They periodically select new test samples for malware analysts to label, then add these labeled samples to the training set and retrain the classifier. Analysts have limited bandwidth to label samples every day, and the goal is to make the most efficient use of the analysts' time. There are many schemes for selecting which samples to label; one strong baseline is to select samples where the classifier is most uncertain.

\begin{figure}[t!]
    \centering
    \includegraphics[width=0.47\textwidth]{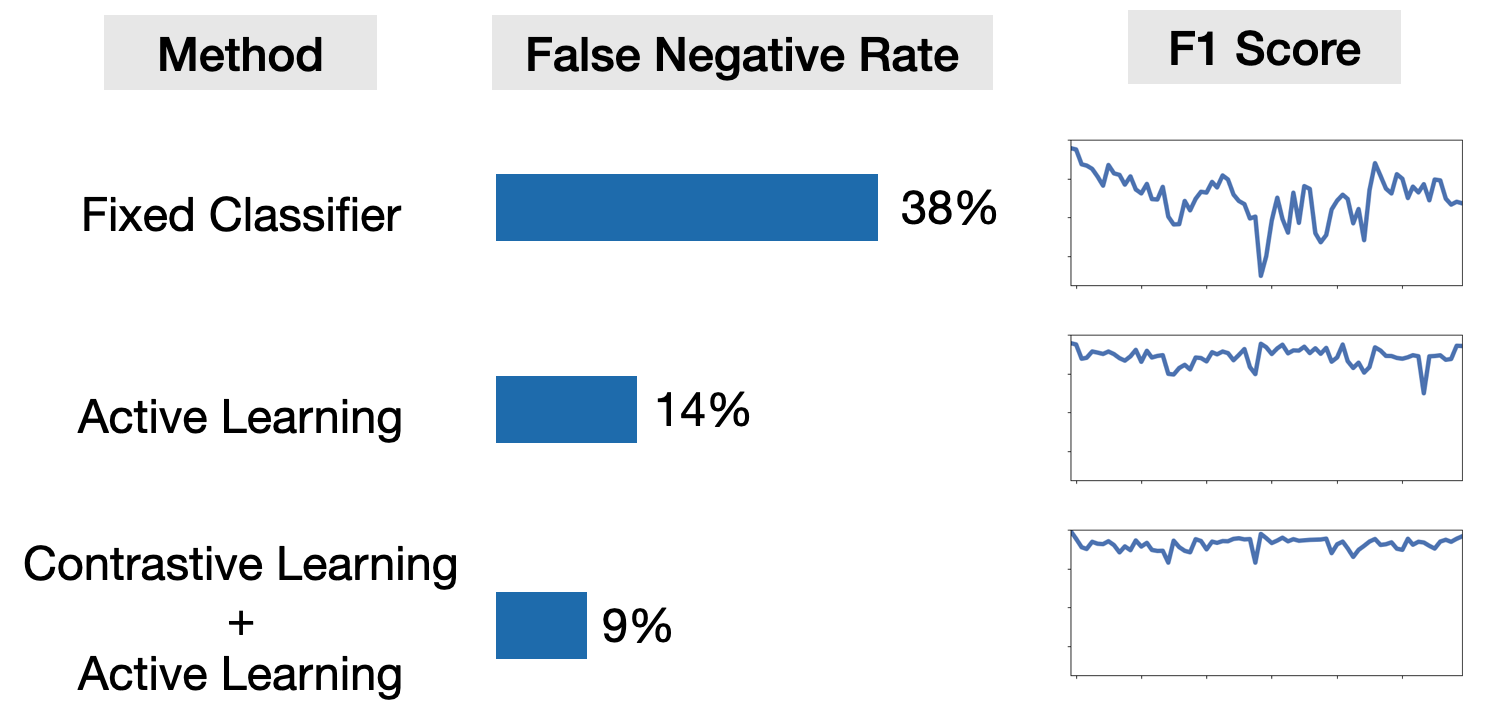}
    \caption{Android malware classification needs a way to update the classifier periodically, to address concept drift.  Training a single classifier is ineffective: the average false negative rate over a 7-year period is 38\%.  State-of-the-art methods for active learning can reduce this to 14\%.  Our method, based on contrastive learning, further improves on past work, reducing the average FNR to 9\% (middle column) and achieving more consistent accuracy over a 7-year period (right column).}
    \label{fig:result-highlight}
\end{figure}

In this paper, we propose a new method of active learning for Android malware detection.
Our goal is to reduce the amount of human analyst effort needed to achieve a fixed performance, or improve classifier performance given a fixed level of analyst effort.
Our approach is based on a combination of contrastive learning and a novel method for measuring the uncertainty of such models.
In slogan form, we propose that continuous learning for security tasks is enabled by (hierarchical) contrastive learning plus end-to-end measures of uncertainty.

We show that contrastive learning is well-suited for dealing with concept drift in our dataset.
Figure~\ref{fig:result-highlight} summarizes our results: active learning is necessary to deal with concept drift, and our methods improve on past state-of-the-start schemes, reducing the false negative rate from 14\% to 9\% and ensuring more stable performance of the classifier.

We hypothesize that contrastive learning is well-suited to security tasks because it provides a way to measure similarity of samples.
In contrastive learning, we learn an encoder where similar samples are mapped to nearby vectors in the embedding space, so we can measure the similarity of two samples by calculating the distance between their two embeddings.
When a new malware family or new benign application emerges, we expect it will be dissimilar to all prior samples, hence an appropriate uncertainty measure can recognize that its classification is uncertain and we should have humans analysts label it.
When a malware app experiences gradual drift, or a benign app receives small updates, we can expect that new samples will be similar to past samples and hence a classifier that uses the output of the contrastive encoder may automatically adapt to gradual drift (as the input to the classifier doesn't change much), yielding an architecture that is robust to gradual drift.
Recent work provides evidence that, for image classification, contrastive learning improves robustness against distribution shift~\cite{zhong2022self}.
We provide evidence in this paper that contrastive learning is a good fit for security tasks as well.

%Our key insight is, \emph{similarity-based uncertainty} is more robust against concept drift,
%and it also helps us select better samples to improve the performance of the classifier.
%Contrastive learning is a promising tool
%to summarize similarities between samples.
%Contrastive learning pushes similar samples to be close
%together, and dissimilar samples to be far apart in the embedding space.
%Recently, researchers have shown that
%contrastive learning is more robust than supervised learning against
%data distribution shift~\cite{zhong2022self}.
%Thus, our vision is to combine contrastive learning
%with active learning to continuously detect Android malware.

Security applications pose two unique challenges for contrastive learning that have not been explored before:
detecting new threats while dealing with class imbalance, and measuring uncertainty.

First, in security applications, new threats emerge from time to time, which we must detect and learn to classify correctly.
Also, security applications
exhibit severe class imbalance: in real-world scenarios, most apps are benign
(for instance, 94\% of Android apps in the AndroZoo dataset~\cite{androzoo} are benign).
%and there is a long-tail distribution of sample number in malware families.
%Recently, researchers have shown that contrastive learning can help us compute
%distance-based sample similarity in the embedding space~\cite{yang2021cade}.
We are inspired by CADE~\cite{yang2021cade}, which showed that contrastive learning is promising for detecting new threats (specifically, new malware families).
However, when we experimented with CADE on realistic datasets with class imbalance matching real-world scenarios,
we found that CADE struggles to detect new malware families, often misclassifying them as benign.
To address this, we propose using \emph{hierarchical contrastive learning}.
% a technique previously introduced in the computer vision literature.
Hierarchical contrastive learning allows us to capture the intuition that two malicious samples from the same malware family should be considered very similar; and two malicious samples from different malware families can be considered weakly similar.
% In comparison, contrastive learning treats every pair of malicious training samples as equally similar, regardless of their families.
In comparison, non-hierarchical contrastive learning treats pairs of malicious training samples as dissimilar if they are from different families, and pairs of malicious and benign samples as equally dissimilar.
Thus, hierarchical contrastive learning allows us to take advantage of the additional information that different malware families are weakly similar.
Thereby, hierarchical contrastive learning can more accurately capture that unseen new malware families are more similar to malicious samples than benign samples. 
% even in the presence of heavy class imbalance.

%The difference between hierarchical contrastive learning and regular contrastive learning
%is the following. Regular contrastive learning~\cite{yang2021cade} considers pairs of malware samples
%from the same family to be similar, and pairs of samples from different families
%to be dissimilar (a family is either a malware family or the benign label).
%In comparison, our hierarchical contrastive learning scheme
%has two levels of similarity. %The low level considers pairs of samples from the same
%malware family to be similar. The high level consider
%pairs of arbitrary malware samples to be similar, but not as similar as the low level,
%regardless of whether they are in the same family;
%also, this level considers pairs of benign samples to be similar, and pairs
%of benign and malicious samples to be dissimilar. Using hierarchical contrastive learning,
%we are able to better capture the the notion of benign vs malicious,
%so new malware families will be more similar to malicious samples than benign samples.
%Figure~\ref{fig:motivation} in Section~\ref{sec:hirarchical-contrastive-learning} further explains this idea.

Second, there is no existing measure of uncertainty for a model trained
with contrastive learning.
Standard models map a single sample to a predicted classification, so there are ways to measure the certainty of this prediction.
In comparison, with contrastive learning, training involves a pair of similar or dissimilar samples, so there is no obvious way to assign uncertainty to a single sample.
To solve this problem, we introduce a new uncertainty measure for contrastive learning, which we call \emph{pseudo loss}.
Concretely, given a test sample $x$, we use the classifier to predict the label of $x$.
Then, we construct many pairs of samples that include $x$ and another training sample, compute the contrastive loss on each pair, and average these losses.
A higher average loss value means the model is more uncertain about $x$.
Our active learning scheme then uses this uncertainty measure to select samples with a high uncertainty score for human labelling.

%What's special about our method? Hierarchical contrastive learning.
%(End-to-end training of encoder and classifier.)

%Related work of active learning for malware detection adopts the cold start option.

Third, we identify several engineering improvements that are unique to continuous learning for security.
Active learning can use either cold state learning (where we train a new model from scratch each time) or warm start learning (where we take an older model and continue training it with new samples).
Past work has made little distinction between these two approaches, perhaps because they perform about the same for image classification.
However, we found in our experiments that warm start can offer significant improvements for security classification, when using deep learning.
We suspect this is due to sample imbalance, where in malware detection we typically have a large volume of old labelled samples but few new labelled samples.
Warm start addresses this sample imbalance issue by focusing more on the newest samples.

To evaluate our approach, we collect the APIGraph dataset~\cite{zhang2020enhancing} spanning across seven years from 2012 to 2018, and a new AndroZoo dataset~\cite{androzoo} from 2019 to 2021.
On the APIGraph dataset, we train an initial model using data from 2012.
Then, every month, human analysts label a fixed set of new samples, we expand
the training set, and we update the classifier. We evaluate the performance of this classifier on the next month.
If human analysts label 200 samples each month, 
our approach reduces the false negative rate from 14\% to 9\% (see Figure~\ref{fig:result-highlight}), while also reducing the false positive rate (from 0.86\% to 0.48\%).
As another comparison, if we wish to maintain the same performance of the classifier, our scheme reduces the labelling effort from analysts by 8$\times$ compared to prior methods. On the AndroZoo dataset, the improvement of F1 score ranges from 8.99\% to 16.50\% across different labeling budgets compared to the best prior method.
%We also tried to improve the baseline techniques using our new ideas. Compared to
%the best improved baseline, our approach still reduces the false negative rate by 5\% on average
%across different labeling budgets.

% This is no longer true
% Our method identifies 68\% of new families for human labelling in advance before they become popular, compared to 50\% for the best prior method.
% On average, we label new families 3 months earlier than the best prior method labels them.
% This allows our model to avoid sudden spikes of false negatives when new families emerge, which prior methods struggle with.

Our case study reveals one reason why our scheme performs better: our sample selection method does a better job of identifying new samples for analysts to label. For example, our method identifies samples from the malware family that caused the most false negatives and labels them; the baseline method does not.
This allows our model to quickly recover from a sudden increase of false negatives and avoid future spikes, which prior methods struggle with.

The contribution of this paper is to develop methods for continuous learning for classifying Android malware.
In particular, we evaluate many previously proposed schemes and introduce a new approach that improves significantly on past work in this space.
Borrowing from past work, we show that hierarchical contrastive learning can help address the concept drift problem in malware classification.
We also introduce a novel uncertainty score and method for sample selection, the pseudo loss (Section~\ref{sec:pseudo-loss});
this is the first method we are aware of for measuring uncertainty for a contrastively learned encoder.
We also highlight several engineering lessons (Section~\ref{sec:engineering-lessons}) and show that, in one setting, we can reduce the labeling effort for analysts by $8 \times$. Our code is available at \url{https://github.com/wagner-group/active-learning}.

%To summarize, we make the following contributions in this paper:
%
%\begin{itemize}
%\item We propose a new hierarchical contrastive learning method for continuous Android malware detection.
%\item We propose a new sample selection technique when using contrastive methods with active learning.
%\item We summarize engineering lessons learned for deep active learning.
%\item Our results show that we can reduce the analysts labeling effort by 8$\times$.
%\end{itemize}

\section{Background and Related Work}

\paragraph{Active Learning.}
Many active learning schemes have been proposed in the literature for image and text classification~\cite{settles2009active,ren2021survey,schroder2020survey,liu2022survey,zhan2022comparative}. There are many ways to select samples and update models for active learning. In comparison, relatively few previous works have studied active learning for malware detection~\cite{yang2021bodmas,zhang2020enhancing,yang2021cade}. In our experience, uncertainty sampling is a strong baseline that is hard to beat for malware detection.

\paragraph{OOD Detection.} We focus on the active learning problem in this paper, which needs a sample selection method for continuous learning. Selecting OOD samples is one way to do sample selection.
For instance, uncertainty sampling selects samples with the highest uncertainty score, which can be viewed as a measure of how OOD each sample is.
The prediction confidence of a classifier is commonly used
to detect OOD samples~\cite{li2021can}, and researchers have proposed various methods to calibrate the model's prediction confidence~\cite{guo2017calibration,deo2016prescience}. \textsc{Transcendent} builds on conformal prediction theory~\cite{shafer2008tutorial} to detect OOD samples. \textsc{Transcendent}~\cite{barbero2022transcending,jordaney2017transcend} uses two metrics, credibility and confidence, both utilizing the nonconformity measure to reject test samples that may have drifted. The paper did not provide a way to use the two metrics for active learning.
We extend \textsc{Transcendent} to an active learning scheme by using its metrics to select samples for labeling (Section~\ref{sec:adapted-baseline-setup}) and compare this to our scheme.
CADE~\cite{yang2021cade} uses supervised contrastive learning and a distance-based
OOD score to detect OOD samples. In the paper, the authors have provided a way to use CADE OOD score for retraining a binary SVM classifier. Therefore, we follow the exact same setup as one of the baseline methods in our experiments.
Moreover, we use new ideas to improve CADE for deep active learning and compare our technique against the improved versions.
Previous works have also proposed methods to estimate
uncertainty for neural networks, including Monte-Carlo dropout~\cite{gal2016dropout},
variance of predictions made by a deep ensemble~\cite{lakshminarayanan2017simple},
energy score~\cite{liu2020energy}, focal loss~\cite{mukhoti2020calibrating},
and distance to the k-th nearest neighbor
in the training set~\cite{sun2022out}. OpenOOD~\cite{yangopenood} shows that the detection performance
of different methods vary across different OOD datasets.
Instead of evaluating the detection accuracy on OOD datasets,
we are interested in using uncertainty measures to select samples for active learning,
in order to improve the performance of the classifier.
Researchers have also proposed hierarchical novelty detection by combining hierarchical classification with OOD detection~\cite{lee2018hierarchical}. However, they don't provide an OOD score so we cannot adapt it for active learning. Open set recognition~\cite{geng2020recent,parkmeta} is not helpful in our setting because we need to always predict a binary label (malicious or benign).

\paragraph{Contrastive Learning.}
% SimCLR~\cite{chen2020simple} and MoCo~\cite{chen2020simple}
Contrastive learning is a type of self-supervised learning method that does not
require labels for individual inputs. The only information required is similar and
dissimilar pairs of samples, i.e., the positive pairs and negative pairs. In image applications, we can
use data augmentation over each input image to generate positive pairs, and
consider the rest as negative pairs.
Unsupervised contrastive learning has been proposed for OOD detection~\cite{winkens2020contrastive} in the image domain, but it requires data augmentation techniques that are not available for malware detection.
In this paper, we use supervised contrastive learning~\cite{khosla2020supervised,yang2021cade},
where information about positive and negative pairs come from ground truth
malware family and benign labels.
We are inspired by the promising results from CADE~\cite{yang2021cade} on using supervised contrastive learning to detect drifted samples in Android malware datasets. However, CADE did not experiment with real-world distributions of benign apps. We find that when the majority of data is benign, CADE struggles to detect new malware families as drifted samples. Our new hierarchical contrastive learning scheme can mitigate the class imbalance issue.

Common contrastive learning loss functions include distance-based loss for pairs~\cite{hadsell2006dimensionality,yang2021cade}, triplet loss~\cite{schroff2015facenet},
and normalized cross-entropy loss~\cite{chen2020simple,he2020momentum}.
We build on these ideas to design our loss function for hierarchical contrastive learning.
Hierarchical contrastive learning in the image domain
combines clustering with contrastive learning.
Related papers contrast between cluster assignments~\cite{caron2020unsupervised,doersch2015unsupervised},
contrast between sample and different cluster centroids~\cite{li2020prototypical,wang2021unsupervised},
or select negative samples with probability proportional to dissimilarity of clusters~\cite{guo2022hcsc}.
In comparison, our method does not require any clustering procedure.
The novelty of our work is that we show evidence about what techniques are effective for malware classification, and we improve significantly on past work in this space. Also, our pseudo loss (Section~\ref{sec:pseudo-loss}), used for uncertainty estimation and sample selection, has not appeared in any prior work. Prior work for uncertainty estimation focuses on classification. Our scheme is the first we are aware of for measuring the uncertainty in contrastively learned encoders.

\paragraph{Continous Learning in Malware Detection.}

Previous works have demonstrated the importance of evaluating malware detection on future data that has not been trained on~\cite{allix2015your,miller2016reviewer,pendlebury2019tesseract,arp2022and}.
BODMAS~\cite{yang2021bodmas} compared the following active learning
sample selection schemes: random, uncertain, and non-conformity score~\cite{jordaney2017transcend},
when they are used for PE malware detection. The authors found that
uncertainty sampling performs really well, and the non-conformity score performs
very similar to uncertainty sampling.

Some papers propose better features.
APIGraph~\cite{zhang2020enhancing} proposed to merge semantically similar
features as meta-features for Android malware detection, which can be used on top of
any active learning scheme. The paper did not propose any new active learning method,
and used uncertainty exampling for Android malware detection.
MaMaDroid~\cite{mariconti2016mamadroid} uses sequences of API calls to model
app behaviors, and argues that their features and models require less frequent retraining over time.

Some papers propose online learning methods. DroidOL~\cite{narayanan2016adaptive} and Casandra~\cite{narayanan2017context}
are both online learning methods that continuously train the model after each
new Android app is labeled. This is more expensive compared to training
a model once after labeling a batch of samples in our active learning setting.
DroidEvolver~\cite{xu2019droidevolver} is an online learning method that uses the
classifier to generate pseudo labels for new Android apps, without relying on human labels
in order to avoid manual labeling effort. It was later found that this process
quickly causes the classifier to poison itself~\cite{kan2021investigating}.

Rahman et al.~\cite{rahman2022limitations} have studied continual learning methods for malware detection given storage and training limitations. These methods need to retire old training samples while avoiding catastrophic forgetting, since an antivirus vendor could receive hundreds of thousands of new samples per day. Retiring training samples is out of scope in this paper.

%DroidOL [39] and Casandra [38]

%CADE. Transcending transcend.

%\cite{li2021can} studies whether the prediction uncertainty can be useful to detect drift of Android malware classification. The result says in some situation this helps detect drift. What are these situations?

%\cite{grosse2017adversarial} DREBIN features + MLP.

%\cite{kim2018multimodal} six different DNNs to take static features from manifest file, Dex file, and shared libraries, such as API calls and OPCode frequencies. The features vector is based on similarity with malware representative centroids. MLP.

%\cite{mclaughlin2017deep} disassemble APK files into Smali, extract opcode sequence, and discard operands. CNN.

%\cite{ma2020droidetec} statically extract API call sequences by following the call graphs, and then embed API sequences into vectors. LSTM (RNN).

\section{Methodology}

\begin{figure}[t!]
    \centering
    \includegraphics[width=0.47\textwidth]{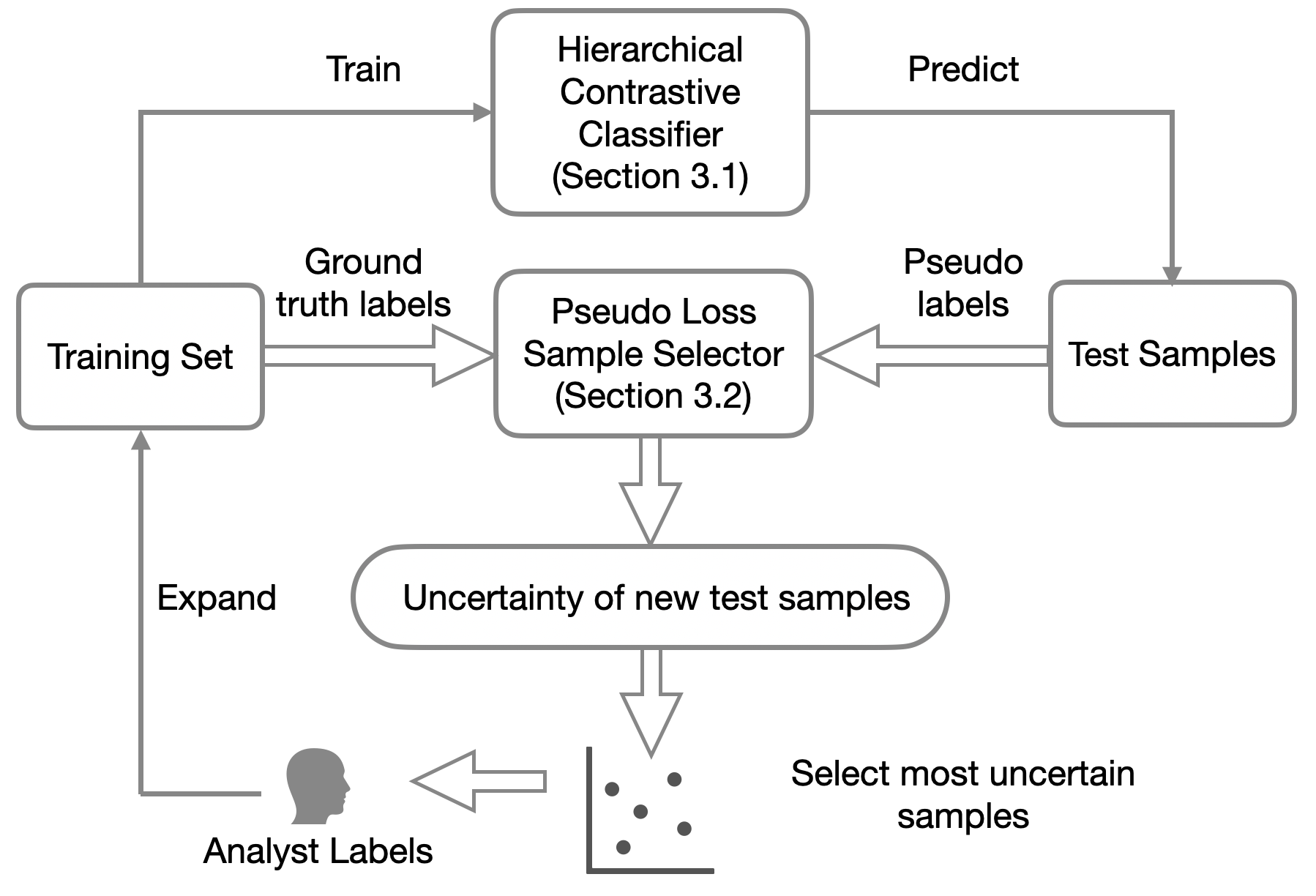}
    \caption{Our approach to continuous learning.}
    \label{fig:workflow}
\end{figure}

\begin{figure}[t]
    \centering
    \includegraphics[width=0.47\textwidth]{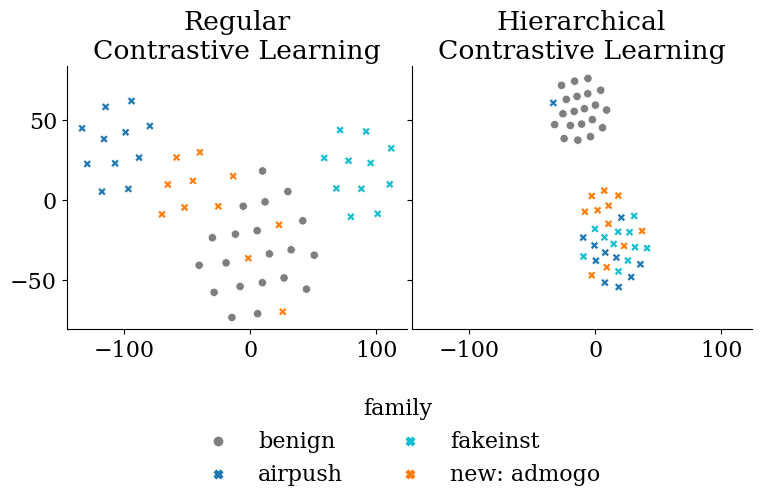}
    \caption{Contrastive learning (left plot) too often treats new malware families as similar to benign samples.  We show a contrastive encoder trained on \texttt{airpush} and \texttt{fakeinst} malware and benign samples, with embeddings visualized using T-SNE.  When the new malware family \texttt{admogo} appears, the contrastive encoder (left) maps many \texttt{admogo} samples (orange x's) near benign samples (gray dots).  Hierarchical contrastive learning (right) does better: the \texttt{admogo} samples now are treated as similar to other malware, even though the model was never trained on any \texttt{admogo} sample.}
    \label{fig:motivation}
\end{figure}

Figure~\ref{fig:workflow} shows our continuous learning framework.
The outer thin arrows in Figure~\ref{fig:workflow} represent the active learning loop.
We continuously expand the training set, train the classifier, and predict the labels of new incoming test samples.
We use hierarchical contrastive learning, which learns an encoder so that similar samples are mapped to nearby embeddings (see Section~\ref{sec:hirarchical-contrastive-learning}).
%The encoder tries to arrange that two samples from the same malware family are mapped to nearby embeddings.

During operation, our scheme repeatedly selects new samples for a human analyst to label.
The inner thick arrows in Figure~\ref{fig:workflow} represent
our new sample selection scheme.
At test time, we compute an uncertainty score for each test sample, based on the predicted label for that test sample and ground truth labels for the training samples.
Then, the sample selector (see Section~\ref{sec:pseudo-loss}) picks the most uncertain samples for the analyst to label.
We assume the human analyst provides both a benign/malware label and a family label for each selected sample.
After we obtain labels for these samples, we  update the model with
contrastive learning to improve the embedding space.
In each iteration, we repeat these steps, to predict labels, measure uncertainty, and update our model.

\subsection{Hierarchical Contrastive Learning}
\label{sec:hirarchical-contrastive-learning}

\subsubsection{Motivating Example}

One of the key challenges of applying contrastive learning to real-world
malware datasets is the data imbalance issue. The majority of samples
are benign, and a contrastively learned model is  likely to
consider an unknown malware sample as similar to benign samples.

The left side of Figure~\ref{fig:motivation} demonstrates this issue.
We trained an autoencoder using a distance-based
contrastive loss function and autoencoder reconstruction loss,
following CADE~\cite{yang2021cade}.
We consider samples with the exact same label as positive pairs,
where each label is a malware family or the benign class.
We consider samples from different labels as negative pairs.
After training, the contrastive autoencoder struggles to separate
new families and benign samples. On the left side of Figure~\ref{fig:motivation},
we plotted the T-SNE visualization of the embeddings for benign samples
and three malware families. Two of the malware families are known and trained on:
\texttt{airpush} and \texttt{fakeinst},
and the other one is a new family \texttt{admogo}.
Regular contrastive learning puts half of the new family samples
inside or nearby the benign region. This behavior makes it hard
for classifiers to accurately detect new families.

We propose hierarchical contrastive learning
to fix this problem. The right side of Figure~\ref{fig:motivation} shows that,
using hierarchical contrastive learning, we can learn an embedding space
that preserves similarity between malicious samples.
We can see that all malware samples fall into a single cluster, and samples from the new family \texttt{admogo} are mapped into this cluster even though this family does not appear in the training set.
Moreover, hierchical contrastive learning also pushes benign and malicious samples
further apart, compared to regular contrastive learning.
We describe how we achieved this in the next section.

\subsubsection{Hierarchical Contrastive Classifier}
\label{sec:hierarchical-contrastive-classifier}

%We use the following notations to describe our training loss function.
We train a hierarchical contrastive classifier $f$ to predict malware.
Our model is separated to two subnetworks. The first subnetwork is an encoder $enc$,
which generates the embeddings for the input $z = enc(x)$.
The second subnetwork acts as the classifier $g$,
which takes the embedding $z$ and predicts a binary label $g(z)$ for the input.

%The model is a concatenation of the encoder and the classifier.

Let $x$ be a sample. The ground truth binary label is $y \in \{0, 1\}$,
where $y = 0$ indicates a benign app, and $y = 1$ indicates a malicious app.
The ground truth multi-class family label is $y'$.
When $y' = 0$, the multi-class label is benign, but otherwise,
it is a malware family label.
Let $f(x) = g(enc(x))$ denote the output for class $y=1$ from
the softmax layer on input $x$; the benign softmax output is $1-f(x)$.
The predicted binary label $\hat{y}$ is $\hat{y}=1$ if $f(x) \ge 0.5$, or $\hat{y}=0$ otherwise.

Intuitively, we construct a loss function that encourages $f(x)$ to correctly predict the label $y$, and also that encourages the encoder to satisfy the following three properties:
\begin{itemize}[leftmargin=*]
    \item If $x_1,x_2$ are two benign samples, or two malicious samples not in the same malware family, then their embeddings should be similar: specifically $\|enc(x_1) - enc(x_2)\|_2 \le m$.
    \item If $x_1,x_2$ are two malicious samples from the same malware family, then their embeddings should be very similar: specifically, $\|enc(x_1) - enc(x_2)\|_2$ should be as small as possible.
    \item If one of $x_1,x_2$ is malicious and the other is benign, then their embedding should be highly dissimilar: specifically $\|enc(x_1) - enc(x_2)\|_2 \ge 2m$.
\end{itemize}
This should hopefully cause benign samples to cluster together, and malicious samples to cluster together; the latter cluster should be composed of many sub-clusters, one for each malware family.
Hopefully, this will encourage the encoder to find invariant properties of malware and of each malware family, and then the classifier will naturally become robust to small shifts in the data distribution.

To achieve this, the training loss is the sum of a  hierarchical contrastive loss and
a classification loss, and we train our model end-to-end with this loss.
Specifically,
\begin{equation}
\mathcal{L} = \mathcal{L}_{hc} + \lambda \mathcal{L}_{ce}
\label{eq:training_loss}
\end{equation}
where $\mathcal{L}_{ce}$ is the classification loss and $\mathcal{L}_{hc}$ is the hierarchical contrastive loss (defined below).
As a common heuristic in machine learning, we choose $\lambda$ such that the average of the two terms
$\mathcal{L}_{hc}$ and $\lambda \mathcal{L}_{ce}$ have a similar mean, so the overall loss is not dominated/overwhelmed by just one term.
The classification loss uses the binary cross entropy loss:
\begin{equation}
\mathcal{L}_{ce} = \sum_{i}\mathcal{L}_{ce}(x_i, y_i)
\label{eq:batch-ce}
\end{equation}
\begin{equation}
\mathcal{L}_{ce}(x_i, y_i) = - y_i\log f(x_i) - (1-y_i) \log (1 - f(x_i))
\label{eq:instance-ce}
\end{equation}
where $i$ ranges over indices of samples in the batch.

The hierarchical contrastive loss computes a loss over pairs of samples in a batch of size $2N$.
%We construct samples with batch size $2N$,
%Let $I$ denote the indices of all samples in the batch.
%The batch is
%divided by two shares.
The first $N$ samples in the batch,
$\{x_k, y_k, y'_k\}_{k=1 \ldots N}$, are sampled randomly. Then, we randomly sample $N$ more samples
such that they have the same label distribution as the first $N$,
i.e., $\{x_{k+N}, y_{k+N}, y'_{k+N}\}_{k=1 \ldots N}$ chosen so that $y_{k} = y_{k+N}$
and $y'_k = y'_{k+N}$.
Let $i$ denote the index of an arbitrary sample within a batch of $2N$ samples.
There are three kinds of pairs in a batch.
We define the following sets to capture them:
\begin{enumerate}[label=\roman*)]
\item Both samples are benign; or, both samples are malicious, but not in the same family. \\ $P(i, y_i, y'_i) \equiv \{j \mid y_j = y_i, y_i = 1 \implies y'_j \ne y'_i, j\neq i\}$
%\item Both samples are benign. \\ $P_b(i, y_i) \equiv \{j \mid y_j = y_i, y_i = 0, j\neq i\}$
%\item Both samples are malicious, but not in the same family. \\ $P_m(i, y_i) \equiv \{j \mid y_j = y_i, y_i = 1, j\neq i\}$
\item Both samples are in the same malware family. \\ $P_z(i, y_i, y'_i) \equiv \{j \mid y'_j = y'_i, y_i = y_j = 1, j\neq i\}$
\item One sample is benign and the other is malicious. \\ $N(i, y_i) \equiv \{j \mid y_j \neq y_i\}$
\end{enumerate}
%We can merge the first two sets into a positive pairs set
%$P(i, y_i) \equiv \{j \mid y_j = y_i, j\neq i\} = P_b(i, y_i) \cup P_m(i, y_i)$.
These sets capture multiple degrees of similarity: $P(i, y_i, y'_i)$ contains pairs that are considered weakly similar, $P_z(i, y_i, y'_i)$ contains pairs that are highly similar, and $N(i, y_i)$ pairs that are dissimilar.

Let $d_{ij}$ denote the euclidean distance between two arbitrary samples $i$ and $j$ in the embedding space:
$d_{ij} = \lVert enc(x_i) - enc(x_j) \rVert_2$.
Let $m$ denote a fixed margin (a hyperparameter).
The hierarchical contrastive loss is defined as:
\begin{equation}
\mathcal{L}_{hc} = \sum_{i}\mathcal{L}_{hc}(i)
\label{eq:lhc}
\end{equation}
\begin{align}
\begin{split}
\mathcal{L}_{hc}(i) = & \frac{1}{|P(i, y_i, y'_i)|}\sum_{j \in P(i, y_i, y'_i)}\max(0, d_{ij} - m) \\
 + & \frac{1}{|P_z(i, y_i, y'_i)|}\sum_{j \in P_z(i, y_i, y'_i)}d_{ij} \\
 + &  \frac{1}{|N(i, y_i)|}\sum_{j \in N(i, y_i)}\max(0, 2m - d_{ij})
\end{split}
\label{eq:instance-contrastive-loss}
\end{align}
%The hierarchical contrastive loss is averaged across all samples $i$
%in a batch, in Equation~\eqref{eq:lhc}.
%Specifically, Equation~\eqref{eq:instance-contrastive-loss} describes
%the loss for each sample.
The hierarchical contrastive loss has three terms.
The first term asks positive pairs from $P(i, y_i, y'_i)$
to be close together, but we don't require them to be too close.
We only penalize the distance between these pairs
if it is larger than $m$. Specifically, these are (benign, benign)
and (malicious, malicious) pairs. This term is helpful for us to
learn properties that are common to all malicious apps or all benign apps.
The second term asks samples from the same
malware family to be treated as very similar, and we penalize any non-zero distance $d_{ij}$ between them.
The last term aims to separate benign and malicious samples from each other, hopefully at least $2m$ apart from each other; if the distance is already larger than $2m$, we don't care how far apart they might be.

\subsection{Pseudo Loss Sample Selector}
\label{sec:pseudo-loss}

Next, we introduce a novel way to compute an uncertainty score for a test sample, for a hierarchical contrastive classifier.
This score is used in active learning: the samples with the highest uncertainty scores are selected for analysts to label.
We face three challenges:
\begin{enumerate}[label=(\roman*)]
\item We need to take into account the uncertainty of both the encoder and the classifier subnetworks in our model.
\item We need a new way to measure uncertainty for the hierarchical contrastive encoder.
Past work has only considered uncertainty scores for classifiers, but not for contrastive encoders.
\item The uncertainty measure should be efficient to compute.
\end{enumerate}

\subsubsection{Key Idea}
\label{sec:key-idea}

Our design is motivated by an unsupervised learning view
on how researchers measure uncertainty for neural network classifiers.
The basic idea is, if we use the predicted label instead of the ground truth label
to compute the classification loss for an input, the loss value represents
the uncertainty of the classifier. We call this the \emph{pseudo loss},
since we can view the predicted label as a pseudo label for the input and compute the loss with respect to this pseudo label.

For example, a common uncertainty measure for a neural network
is to use one minus the max softmax output of the network.
For our encoder-classifier model, using the notations introduced
in Section~\ref{sec:hierarchical-contrastive-classifier},
the uncertainty score would be:
\begin{equation}
\mathcal{U}(x) = 1 - \max (f(x), 1-f(x)).
\label{eq:max-softmax-unc}
\end{equation}
Alternatively, we can view this as an instance of a pseudo loss.
Let $\hat{y}$ denote the binary label predicted by $f(x)$, i.e., $\hat{y} = 1$ if $f(x) \ge 1-f(x)$ or $\hat{y} = 0$ otherwise.
Then the cross-entropy loss with respect to $\hat{y}$ is given by
\begin{align}
\begin{split}
\mathcal{L}_{ce}(x, \hat{y}) = & - \hat{y} \log f(x) - (1-\hat{y}) \log (1 - f(x)) \\
 = & - \max (\log f(x), \log (1 - f(x))).
\end{split}
\label{eq:pseudo-ce}
\end{align}
Since $\log$ is a monotonic function, combining Equation~\eqref{eq:max-softmax-unc}
and Equation~\eqref{eq:pseudo-ce}, we have
$\mathcal{L}_{ce}(x, \hat{y}) = -\log (1-\mathcal{U}(x))$.
Thus, ranking samples by $\mathcal{U}(x)$ gives the same ranking as $\mathcal{L}_{ce}(x, \hat{y})$.
Therefore, the pseudo loss $\mathcal{L}_{ce}(x, \hat{y})$ is a reasonable uncertainty score, one that is equivalent to the standard softmax confidence uncertainty.

The benefit of the pseudo loss formulation is that it can be applied to any learned model, not just classification.
Therefore, our main insight is that we can derive an uncertainty score for a hierarchical contrastive model by constructing a pseudo loss from the training loss defined in Equation~\eqref{eq:training_loss}. 

\subsubsection{Pseudo Loss for Contrastive Learning}

To realize our idea of the pseudo loss for contrastive learning,
there is still a key difference from supervised learning.
The uncertainty of a sample in supervised learning depends on
only the sample, but the uncertainty of the sample in contrastive learning
depends on other samples as well.
Since our goal is to measure uncertainty in a way that reflects the encoder's similarity measure,
we compare the test sample with nearby training samples.

We use the following procedure to compute the pseudo loss for contrastive learning.
Given a test sample $x_i$, we compute its embedding $enc(x_i)$, as well as the embedding of all training samples.\footnote{In our experiments, we normalize the embeddings to have unit length, but in retrospect, we expect normalization is unnecessary.}
Then, we find the $2N-1$ nearest neighbors in the training set to $x_i$, with distances computed in the normalized embedding space.
We obtain a batch of $2N$ samples, containing $x_i$ and its $2N-1$ neighbors.
We use the predicted binary label $\hat{y}_i$ for $x_i$ as a pseudo label for $x_i$, and use the ground truth label for all $2N-1$ training samples.
These labels allow us to compute the positive and negative pairs
in the batch, so we can compute the training loss of the test sample
using Equation~\eqref{eq:instance-contrastive-loss}.

In particular, given a test sample $x_i$, we define the pseudo loss
for hierarchical contrastive learning as:
\begin{align}
\begin{split}
\hat{\mathcal{L}}_{hc}(i) = & \frac{1}{|\hat{P}(i, \hat{y_i})|}\sum_{j \in \hat{P}(i, \hat{y_i})}\max(0, d_{ij} - m) \\
 + &  \frac{1}{|N(i, \hat{y_i})|}\sum_{j \in N(i, \hat{y_i})}\max(0, 2m - d_{ij})
\end{split}
\label{eq:pseudo-loss}
\end{align}

In other words, this uses $\hat{y_i}$ instead of $y_i$ to compute the first and third terms
in $\mathcal{L}_{hc}(i)$ (Equation~\eqref{eq:instance-contrastive-loss}).
Note that since our pseudo label is a binary label, we do not
have multi-class pseudo label information and we cannot compute the second term in Equation~\eqref{eq:instance-contrastive-loss}, so we omit it. Moreoever, for the first term, $\hat{P}(i, \hat{y_i})$ is slightly different from $P(i, y_i, y'_i)$. We define $\hat{P}(i, \hat{y_i})| \equiv \{j \mid y_j = \hat{y_i}, j\neq i\}$.
This yields an uncertainty score that generalizes the prediction uncertainty of a neural network classifier
to contrastive learning.

\subsubsection{Sample Selector}

Putting all of this together, we use the pseudo loss version of the training loss
for our hierarchical contrastive classifier to measure its uncertainty.
Based on Section~\ref{sec:key-idea}, we define the pseudo loss
for binary cross entropy as
\begin{equation}
\hat{\mathcal{L}}_{ce}(i) = \mathcal{L}_{ce}(x_i, \hat{y_i}).
\end{equation}
We measure the uncertainty of our model given an input $x_i$ as
\begin{equation}
\hat{\mathcal{L}}(i) = \hat{\mathcal{L}}_{hc}(i) + \lambda \hat{\mathcal{L}}_{ce}(i).
\label{eq:our-uncertainty}
\end{equation}

This uncertainty score solves all three challenges listed earlier.
It captures both the uncertainty of the encoder
and the uncertainty of the classifier, and it is efficient to compute.
At test time, we use Equation~\eqref{eq:our-uncertainty}
to compute uncertainty scores for all test samples.
Then, we label the samples with the highest uncertainty scores
for active learning.

\begin{figure}[ht!]
    \centering
    \includegraphics[width=0.47\textwidth]{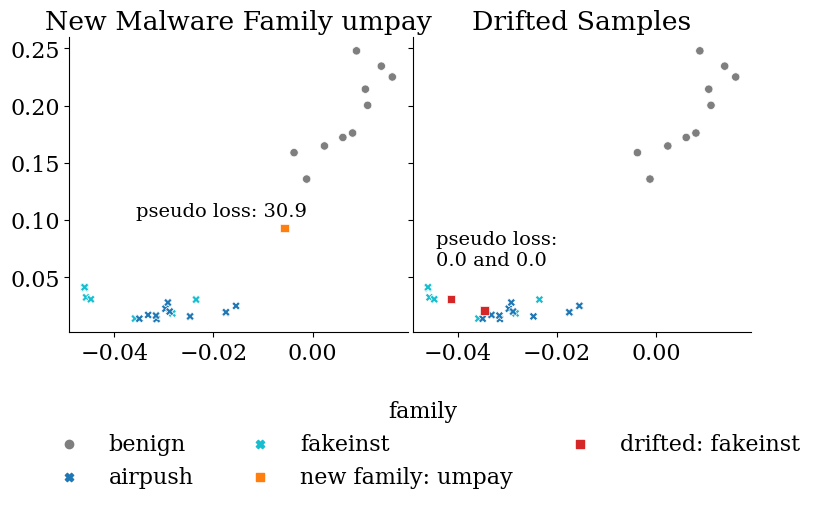}
    \caption{Our pseudo loss uncertainty score is effective at identifying new malware families. We train a contrastive classifier on benign samples and malicious samples from the airpush and fakeinst families.  On the left, we add a test sample from a new malware family; as shown, its pseudo loss uncertainty score is very high, so it would be selected for human labelling.  On the right, we add two test samples from an existing malware family that experienced drift; as shown, their pseudo loss uncertainty scores are very low, indicating that they do not need to be labelled by humans.}
    \label{fig:pseudoloss}
\end{figure}

Figure~\ref{fig:pseudoloss} illustrates our uncertainty score in action.
The left side shows that a sample from a new malware family (\texttt{umpay}) has high pseudo loss, and thus is selected for human labelling.
Intuitively, this sample lies between the benign cluster and malicious cluster,
so its nearest neighbors in the training set contain conflicting labels,
which results in a high loss value for contrastive learning.
%This sample was selected for human labels by our active learning procedure in the experiments.
The right side shows that
two drifted samples from a known malware family (\texttt{fakeinst}) have low pseudo loss, since they are very close to other malicious training samples.
Our active learning procedure does not label them, since they are among samples with the
lowest pseudo loss values. Since the classifier works on embedding vectors, we can expect the classifier to classify them correctly.

\section{Evaluation}

In this section, we compare our new method against two kinds of other schemes:
1) active learning techniques from previously published papers on malware detection,
and 2) improved active learning schemes we tried, inspired by prior work.
We also discuss new lessons learned for applying deep active learning for malware detection.

\subsection{Dataset}

We evaluate on two datasets, from APIGraph~\cite{zhang2020enhancing} and AndroZoo~\cite{androzoo} respectively.

We use the list of app hashes provided by APIGraph~\cite{zhang2020enhancing} 
to collect Android apps spanning over 7 years, from 2012 to 2018. APIGraph uses the appearance timestamps from VirusTotal to order the apps over 7 years.
The way that APIGraph collects the dataset has carefully addressed the spatial bias and temporal bias that commonly exists in malware datasets~\cite{pendlebury2019tesseract,arp2022and}:
90\% of the apps are benign; and the samples are ordered and almost evenly distributed across 7 years that allows time-consistent experiments. Specifically, we collect malware apps from VirusTotal~\cite{virustotal}, VirusShare~\cite{virusshare}, and the AMD dataset~\cite{wei2017deep} and benign apps from AndroZoo~\cite{androzoopaper, androzoo}.
The final number of apps we collected are shown in Table~\ref{tab:apigraph-dataset}.

% for year in $(seq 2012 2018); do find /data2/yizheng/mldroid_drebin/malware/$year/ | grep "\.data$" | wc -l; done
% for year in $(seq 2012 2018); do find /data2/yizheng/mldroid_drebin/benign/$year/ | grep "\.data$" | wc -l; done
\begin{table}[t!]
  \centering
  \small
	\begin{tabular}{c|r|r|r|r}
		\hline
		\textbf{Year} & \begin{tabular}[c]{@{}c@{}}\textbf{Malicious}\\\textbf{Apps}\end{tabular} & \begin{tabular}[c]{@{}c@{}}\textbf{Benign}\\\textbf{Apps}\end{tabular}  & \begin{tabular}[c]{@{}c@{}}\textbf{Total}\end{tabular} & \begin{tabular}[c]{@{}c@{}}\textbf{Malware}\\\textbf{Families}\end{tabular} \\\hline
% generic labels, not counting the "unknown" family
            2012  & 3,061 & 27,472 & 30,533 & 104 \\
		2013  & 4,854 & 43,714 & 48,568 & 172 \\
		2014  & 5,809 & 52,676 & 58,485 & 175 \\
		2015  & 5,508 & 51,944 & 57,452 & 193 \\
		2016  & 5,324 & 50,712 & 56,036 & 199 \\
		2017  & 2,465 & 24,847 & 27,312 & 147 \\
		2018  & 3,783 & 38,146 & 41,929 & 128 \\

% new family FAM: label, some files with .data
		% 2012  & 3,025 & 27,472 & 30,497 & 68 \\
		% 2013  & 4,854 & 43,714 & 48,568 & 132 \\
		% 2014  & 5,809 & 52,676 & 58,485 & 150 \\
		% 2015  & 5,509 & 51,944 & 57,453 & 146 \\
		% 2016  & 5,361 & 50,712 & 56,073 & 155 \\
		% 2017  & 2,466 & 24,847 & 27,313 & 122 \\
		% 2018  & 3,783 & 38,146 & 41,929 & 112 \\
		\hline
	\end{tabular} 
	\caption{We collect Android apps from the APIGraph dataset~\cite{zhang2020enhancing} spanning across seven years. Within total apps, 10\% of them are malicious apps.}
	\label{tab:apigraph-dataset}
\end{table}

In addition, we collect a new dataset of Android apps from AndroZoo~\cite{androzoo} that appeared from 2019 to 2021. We randomly sample malware apps with more than 15 detections by antivirus engines in VirusTotal, and randomly sample benign apps with 0 detection. For each month, the ratio of benign apps to malicious apps is 9:1. Table~\ref{tab:androzoo-dataset} shows the overall statistics of the AndroZoo dataset. In the year of 2021, the available malware apps on AndroZoo is fewer than the previous years.

We query VirusTotal and then use AVClass2~\cite{sebastian2020avclass2} to obtain the family label for malicious apps. If an app does not have any family label\footnote{The output from AVClass2 does not have a family label other than ``Android'' or ``grayware''.}, we use the ``unknown'' family label.

% old numbers
% On average, every month we have 384 malicious apps, 50 families, and 3,832 benign apps. Table~\ref{tab:dataset} also shows the number of malware families seen every year.
% average 163 families every year

We extract DREBIN features~\cite{arp2014drebin} from the apps to train all models. DREBIN uses 8 sets of features to capture the app's required access to hardware components, requested permissions, names of app components, filtered intents, usage of restricted API calls, actually used permissions, suspicious API calls, and network addresses.

% since all samples are at least 5 years old, 
As is typical for research on active learning in malware classification, we simulate the human analyst using post-facto data from VirusTotal and AVClass2.
Our assumption is that over time VirusTotal scores converge to the correct label; we treat current VirusTotal and AVClass2 labels as ground truth, and whenever an active learning scheme calls for a human analyst to label a sample, we use these ground-truth labels.

We apply each active learning scheme to select new samples each month, update/retrain the classifier, and then predict on samples from the next month.

\subsection{Active Learning Setup}
\label{sec:dataset-split}

% \subsubsection{Dataset Split}

We found out that, hyperparameter tuning makes a big difference in the performance of the classifier in the active learning setting. Moreover, for deep active learning schemes including our method, warm start performs better than cold start. Warm start continues training the model from previously learned weights, and cold start retrains the model from scratch. We will summarize engineering lessons learned in Section~\ref{sec:engineering-lessons}.

% removed
% We perform two rounds of hyperparameter tuning.
% The first round aims to find several good ways to train the initial classifier. The second round aims to find the best way to continuously update the initial classifier during active learning.
% \textbf{Randomized data split.} For the first round, we randomly perform five train and validation splits from the 2012 training data. For different hyperparameters to train a model, we compute the average validation performance. Then, we keep the best 10 sets of hyperparameters for the next round.

\textbf{Time-consistent data split.} We choose hyperparameters that perform the best in active learning. We split the data into a training set (the first year of apps), a validation set (the next six months), train an initial classifier on the training set, and then use active learning with a labeling budget of 50 samples per month on the validation set to select the best hyperparameters.
%The validation performance is calculated on the future month that has not been trained on, and averaged across six months.
After finding the best hyperparameters, we test the active learning performance using data from the remaining months. For the APIGraph dataset, the training set is 2012 data, the validation set is 2013-01 to 2013-06, and the test set covers 2013-07 to 2018-12. For the AndroZoo dataset, the training set is 2019 data, the validation set is 2020-01 to 2020-06, and the test set is 2020-07 to 2021-12. The test performance is averaged across all test months. More details about the training samples are in Appendix~\ref{appendix:details-training-samples}.

\begin{table}[t!]
  \centering
  \small
	\begin{tabular}{c|r|r|r|r}
		\hline
		\textbf{Year} & \begin{tabular}[c]{@{}c@{}}\textbf{Malicious}\\\textbf{Apps}\end{tabular} & \begin{tabular}[c]{@{}c@{}}\textbf{Benign}\\\textbf{Apps}\end{tabular}  & \begin{tabular}[c]{@{}c@{}}\textbf{Total}\end{tabular} & \begin{tabular}[c]{@{}c@{}}\textbf{Malware}\\\textbf{Families}\end{tabular} \\\hline
		2019  & 4,542 & 40,947 & 45,489 & 121 \\
		2020  & 3,982 & 34,921 & 38,904 & 82 \\
		2021  & 1,676 & 13,985 & 15,662 & 51 \\
		\hline
	\end{tabular} 
	\caption{We collect a new AndroZoo dataset by randomly sampling malware and benign apps from AndroZoo~\cite{androzoo}. In the dataset, 10\% of all apps are malicious.}
	\label{tab:androzoo-dataset}
\end{table}

\subsection{Comparison against Baselines}
\label{sec:compare-baselines}

\subsubsection{Baseline Active Learning Schemes}
\label{sec:baseline-setup}

\begin{table*}[ht!]
  \centering
  \normalsize
  \begin{tabular}{c | c | c | rrr | rrr }
    \hline
    \multirow{3}{*}{\begin{tabular}{@{}c@{}}{\bf Monthly}\\{\bf Sample}\\{\bf Budget}\end{tabular}}  & \multirow{3}{*}{\begin{tabular}{@{}c@{}}{\bf Model}\\{\bf Architecture}\end{tabular}} &  \multirow{3}{*}{\begin{tabular}{@{}c@{}}{\bf Sample}\\{\bf Selector}\end{tabular}} & \multicolumn{3}{|c}{\begin{tabular}{@{}c@{}}{\bf APIGraph Dataset}\end{tabular}} & \multicolumn{3}{|c}{\begin{tabular}{@{}c@{}}{\bf AndroZoo Dataset}\end{tabular}} \\
    
     &  &  & \multicolumn{3}{|c}{Average Performance (\%)} & \multicolumn{3}{|c}{Average Performance (\%)} \\
     &  &  & FNR & FPR & F1 & FNR & FPR & F1 \\
    \hline
    \hline
    % \multirow{6}{*}{10} &  MLP & Uncertainty &  &  &  &  &  &   \\
    % \cline{2-9}
    % &  \multirow{2}{*}{SVM} & Uncertainty & {\bf } & {\bf } & {\bf }  &  &  &    \\
    % &  &  CADE OOD &  &  &   &  &  &    \\
    % \cline{2-9}
    % & GBDT & Uncertainty &  &  &   &  &  &     \\
    % \cline{2-9}
    % & Ours: & \multirow{2}{*}{Pseudo Loss} & {\bf } & {\bf } & {\bf } &  &  &     \\
    % &  Enc + MLP &  & \green{($\downarrow$ )} & \red{($\uparrow$ )} & \green{($\uparrow$ )}  &   &  &  \\
    % \hline
    % \hline
    \multirow{9}{*}{50} &  Binary MLP & Uncertainty & 23.77 & 0.52 & 83.84 & 53.12 & 0.46 & 59.50  \\
    \cline{2-9}
     &  Multiclass MLP & Uncertainty & 16.10 & 4.64 &  73.77 & 49.86 & 28.52 & 28.65  \\
    \cline{2-9}
     & \multirow{2}{*}{\begin{tabular}{@{}c@{}}{Multiclass MLP}\\{+ Binary SVM}\end{tabular}} & \multirow{2}{*}{Uncertainty} &  \multirow{2}{*}{38.40} & \multirow{2}{*}{1.01} & \multirow{2}{*}{71.38} & \multirow{2}{*}{73.13} & \multirow{2}{*}{2.87} & \multirow{2}{*}{34.04}  \\
    &  &  &  &  &   &  &  &    \\
    \cline{2-9}
    &  \multirow{2}{*}{Binary SVM} & Uncertainty & {\bf 16.92} & {\bf 0.61} & {\bf 87.72}  &  {\bf 48.77}  & {\bf 0.29}  & {\bf 63.42}  \\
    &  &  CADE OOD & 36.11 & 12.9 & 71.70  & 62.01 &  0.55 & 50.26   \\
    \cline{2-9}
     &  Multiclass SVM & Uncertainty & 35.79 & 0.17 & 87.43 & 65.77 & 0.09 & 46.91 \\
    \cline{2-9}
    & Binary GBDT & Uncertainty & 31.75 & 0.54 & 77.92  & 50.35 & 0.47 & 61.06    \\
    \cline{2-9}
    & \multirow{2}{*}{Ours: Enc + MLP} & \multirow{2}{*}{Pseudo Loss} & {\bf 15.15} & {\bf 0.52} & {\bf 89.23} & {\bf 27.65} & {\bf 0.53} & {\bf 79.92}    \\
    &   &  & \green{($\downarrow$ 1.77)} & \green{($\downarrow$ 0.09)} & \green{($\uparrow$ 1.51)}  &  \green{($\downarrow$ 21.12)} & \red{($\uparrow$ 0.24)} & \green{($\uparrow$ 16.50)}  \\
    \hline
    \hline
    \multirow{9}{*}{100} & Binary MLP & Uncertainty & 20.64 & 0.49 & 86.03 & 46.39 & 0.30 & 65.26  \\
    \cline{2-9}
     &  Multiclass MLP & Uncertainty & 14.77 & 6.44 & 69.91 & 35.34 & 32.64 & 33.72  \\
    \cline{2-9}
     &  \multirow{2}{*}{\begin{tabular}{@{}c@{}}{Multiclass MLP}\\{+ Binary SVM}\end{tabular}} & \multirow{2}{*}{Uncertainty} &  \multirow{2}{*}{30.45} & \multirow{2}{*}{1.76} & \multirow{2}{*}{74.11} & \multirow{2}{*}{73.47} & \multirow{2}{*}{3.88} & \multirow{2}{*}{31.69}  \\
     &  &  &  &  &   &  &  &    \\
    \cline{2-9}
    &  \multirow{2}{*}{Binary SVM} & Uncertainty & {\bf 15.41} & {\bf 0.68} & {\bf 88.38}  & {\bf 43.07} & {\bf 0.32} & {\bf 68.33}   \\
    &  &  CADE OOD & 23.48 & 0.96 & 82.22  & 58.78 &  0.70 & 52.47   \\
    \cline{2-9}
     &  Multiclass SVM & Uncertainty & 28.36 & 0.17 & 82.18 & 54.29 & 0.12 & 58.26 \\
    \cline{2-9}
    & Binary GBDT & Uncertainty & 27.76 & 0.67 & 80.15  & 48.59 & 0.76 & 62.58    \\
    \cline{2-9}
    & \multirow{2}{*}{Ours: Enc + MLP} & \multirow{2}{*}{Pseudo Loss} & {\bf 13.69} & {\bf 0.44} & {\bf 90.42} & {\bf 27.35} & {\bf 0.41} & {\bf 80.07}    \\
    &   &  & \green{($\downarrow$ 1.72)} & \green{($\downarrow$ 0.24)} & \green{($\uparrow$ 2.04)}  & \green{($\downarrow$ 15.72)}  & \red{($\uparrow$ 0.09)} & \green{($\uparrow$ 11.74)} \\
    \hline
    \hline
    \multirow{9}{*}{200} & Binary MLP & Uncertainty & 19.71 & 0.39 & 86.97 & 42.57 & 0.34 & 68.47  \\
    \cline{2-9}
     &  Multiclass MLP & Uncertainty & 14.56 & 4.26 & 75.65 & 39.78 & 34.76 & 28.59  \\
    \cline{2-9}
     &  \multirow{2}{*}{\begin{tabular}{@{}c@{}}{Multiclass MLP}\\{+ Binary SVM}\end{tabular}} & \multirow{2}{*}{Uncertainty} &  \multirow{2}{*}{29.46} & \multirow{2}{*}{1.98} & \multirow{2}{*}{74.09} & \multirow{2}{*}{70.32} & \multirow{2}{*}{0.93} & \multirow{2}{*}{39.51}  \\
      &  &  &  &  &   &  &  &    \\
    \cline{2-9}
    &  \multirow{2}{*}{Binary SVM} & Uncertainty & {\bf 14.07} & {\bf 0.86} & {\bf 88.47}  & {\bf 40.31} & {\bf 0.37} & {\bf 70.24}   \\
    &  &  CADE OOD & 21.68 & 0.67 & 84.50  & 51.32 & 0.78 & 59.11   \\
    \cline{2-9}
     &  Multiclass SVM & Uncertainty & 21.19 & 0.21 & 86.90 & 44.77 & 0.13 & 66.55 \\
    \cline{2-9}
    & Binary GBDT & Uncertainty & 24.71 & 0.56 & 82.71  & 42.97 & 0.80 & 67.28    \\
    \cline{2-9}
    & \multirow{2}{*}{Ours: Enc + MLP} & \multirow{2}{*}{Pseudo Loss} & {\bf 9.42} & {\bf 0.48} & {\bf 92.72} & {\bf 27.67} & {\bf 0.39} & {\bf 80.51}    \\
    &   &  & \green{($\downarrow$ 4.65)} & \green{($\downarrow$ 0.38)} & \green{($\uparrow$ 4.25)}  & \green{($\downarrow$ 12.64)}  & \red{($\uparrow$ 0.02)} & \green{($\uparrow$ 10.27)} \\
    \hline
    \hline
    \multirow{9}{*}{400} &  Binary MLP & Uncertainty & {\bf 16.04} & {\bf 0.40} & {\bf 89.25} & 36.25 & 0.34 & 73.70  \\
    \cline{2-9}
     &  Multiclass MLP & Uncertainty & 15.07 & 4.15 & 75.94 & 34.48 & 24.44 & 38.34  \\
    \cline{2-9}
     &  \multirow{2}{*}{\begin{tabular}{@{}c@{}}{Multiclass MLP}\\{+ Binary SVM}\end{tabular}} & \multirow{2}{*}{Uncertainty} &  \multirow{2}{*}{28.85} & \multirow{2}{*}{1.68} & \multirow{2}{*}{75.69} & \multirow{2}{*}{73.94} & \multirow{2}{*}{1.92} & \multirow{2}{*}{33.74}  \\
      &  &  &  &  &   &  &  &    \\
    \cline{2-9}
    &  \multirow{2}{*}{Binary SVM} & Uncertainty & 12.86 & 0.90 & 89.02 & 34.73 & 0.43 & 74.12   \\
    &  &  CADE OOD & 20.61 & 0.59 & 85.52  & 49.98 &  0.94 &  59.53  \\
    \cline{2-9}
     &  Multiclass SVM & Uncertainty & 17.87 & 0.24 & 88.88 & 40.99 & 0.14 & 69.61 \\
    \cline{2-9}
    & Binary GBDT & Uncertainty & 20.16 & 0.46 & 86.24  & {\bf 33.62} & {\bf 0.38} & {\bf 76.82}    \\
    \cline{2-9}
    & \multirow{2}{*}{Ours: Enc + MLP} & \multirow{2}{*}{Pseudo Loss} & {\bf 7.84} & {\bf 0.50} & {\bf 93.50} & {\bf 21.49} & {\bf 0.31} & {\bf 85.81}    \\
    &   &  & \green{($\downarrow$ 8.20)} & \red{($\uparrow$ 0.10)} & \green{($\uparrow$ 4.25)}  & \green{($\downarrow$ 12.13)}  & \green{($\downarrow$ 0.07)} & \green{($\uparrow$ 8.99)} \\
    \hline
  \end{tabular}
  \caption{Given a fixed monthly labeling budget, we compute the average FNR, FPR, and F1 for different baseline active learning techniques and our method. On the APIGraph dataset, we decrease the labeling cost by 8$\times$ to achieve an average F1 score of over 89\%: our method needs 50 samples / month, and binary MLP needs 400 samples / month. On the AndroZoo dataset, our method reduces the FNR by $1.6\times$ on average, while maintaining under 1\% FPR.
  }
  \label{tab:main}
\end{table*}

The first baseline is active learning with uncertainty sampling. We experiment with uncertainty sampling for both binary and multiclass classifiers. The binary classifiers include a fully-connected neural network (NN), a linear SVM, and gradient boosted decision trees (GBDT)~\cite{yang2021bodmas,zhang2020enhancing}.
We normalize the prediction score from the classifier to between 0 and 1 using softmax for NN, sigmoid for SVM, and the logistic function for GBDT.
The multiclass classifiers include MLP and SVM. We also experiment with a ``Multiclass MLP + Binary SVM" classifier: we train a multiclass MLP first, and then take the penultimate layer as embeddings to train a binary SVM. We consider the ``Multiclass MLP + Binary SVM" a binary classifier. 
The uncertainty score is one minus the max prediction score from all classes. For NN, this is equivalent to the max softmax uncertainty measure.

Our second baseline is active learning with a SVM classifier using the CADE OOD score~\cite{yang2021cade}.
As originally proposed, CADE was primarily envisioned as a way to detect drifted samples; they also use the CADE OOD score to perform one round of active learning using a binary SVM, and we apply that in our setting.
CADE trains a contrastive autoencoder, treating pairs of samples from the same family as similar, and pairs from different families as dissimilar. After training, they define the OOD score of a test sample to be the normalized
distance to the nearest known family.
We perform active learning, each month using their OOD score to select the samples with the highest OOD score for human labelling.

% We experiment with MLP with uncertainty sampling, SVM with uncertainty and CADE OOD sampling, and GBDT with uncertainty sampling as baseline methods, as all of these are used by previously published papers on malware detection.

For all baselines, we use cold start for active learning (i.e., each month we retrain the classifier afresh, from scratch), consistent with past work. We follow the procedure described in Section~\ref{sec:dataset-split} to find the best hyperparameters to train MLP, SVM, and GBDT baseline models, with details in Appendix~\ref{appendix:details-baselines}. For our model, we use warm start, with details in Appendix~\ref{appendix:details-our-model}.

\subsubsection{Results}

We evaluate how much our new technique improves the performance of the classifier on future data compared to the baseline methods.
We experiment with a budget for analyst labels of 50, 100, 200, and 400 samples per month.

\begin{figure*}[t!]
	\centering
        \subfloat[FNR of our technique vs SVM without active learning.]
        {\includegraphics[width=0.48\textwidth]{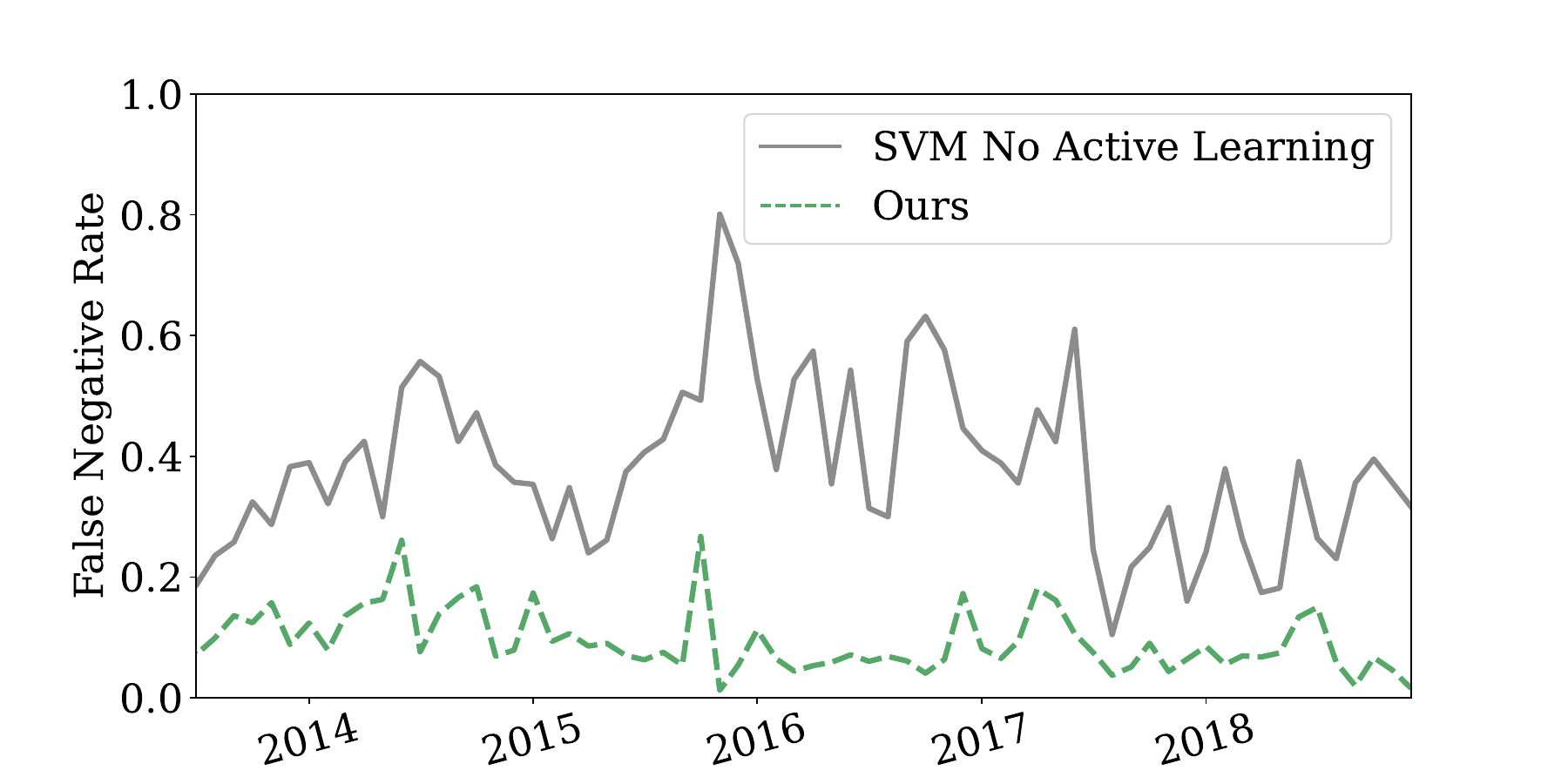}
        \label{fig:ours-plain-cnt200}}
	~
        \subfloat[FNR of our technique vs SVM with uncertainty sampling.]
        {\includegraphics[width=0.48\textwidth]{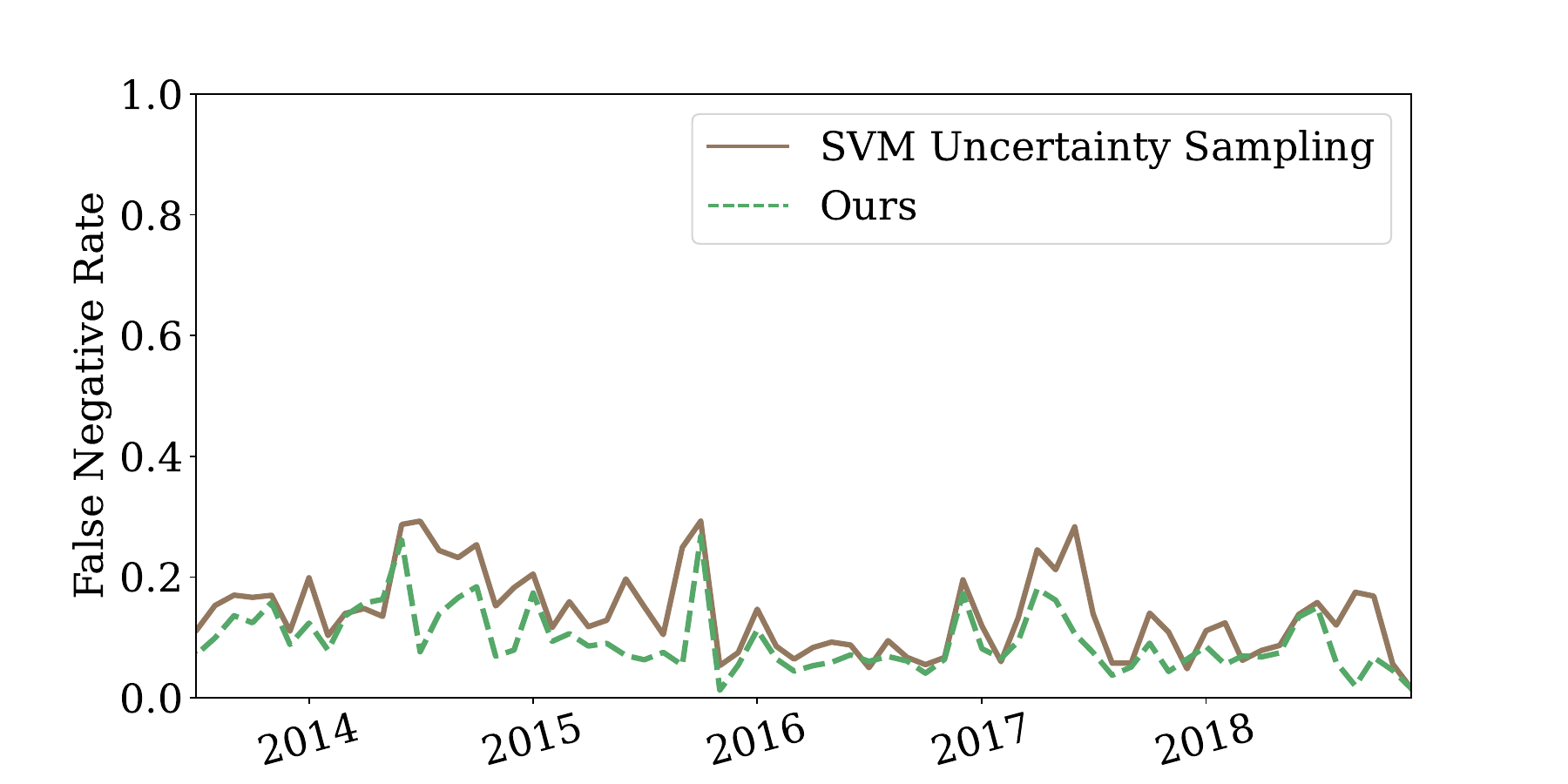}
        \label{fig:ours-svm-unc-cnt200}}

        \subfloat[F1 score of our technique vs SVM without active learning.]
        {\includegraphics[width=0.48\textwidth]{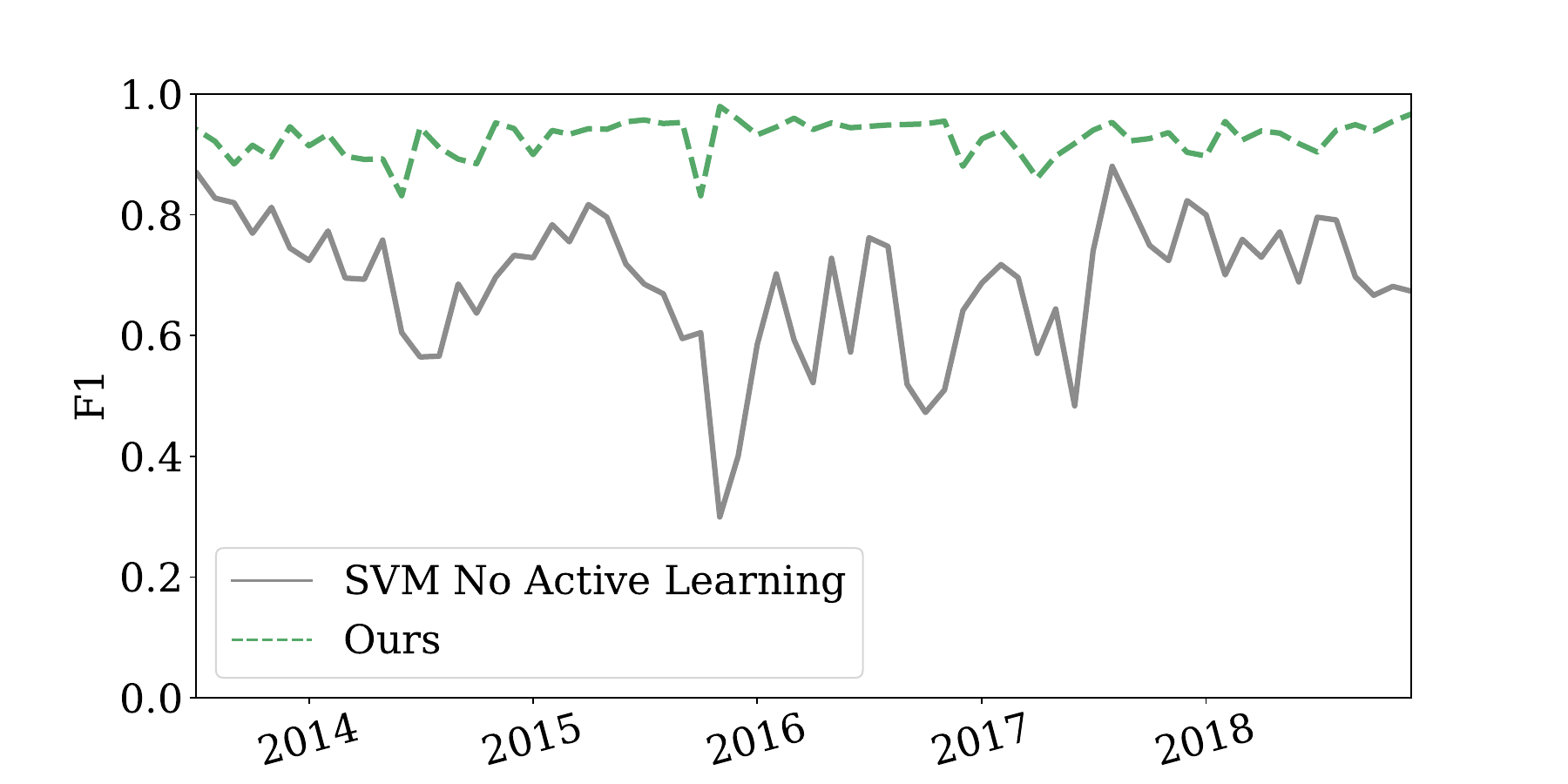}
        \label{fig:ours-plain-cnt200-f1}}
	~
        \subfloat[F1 score of our technique vs SVM with uncertainty sampling.]
        {\includegraphics[width=0.48\textwidth]{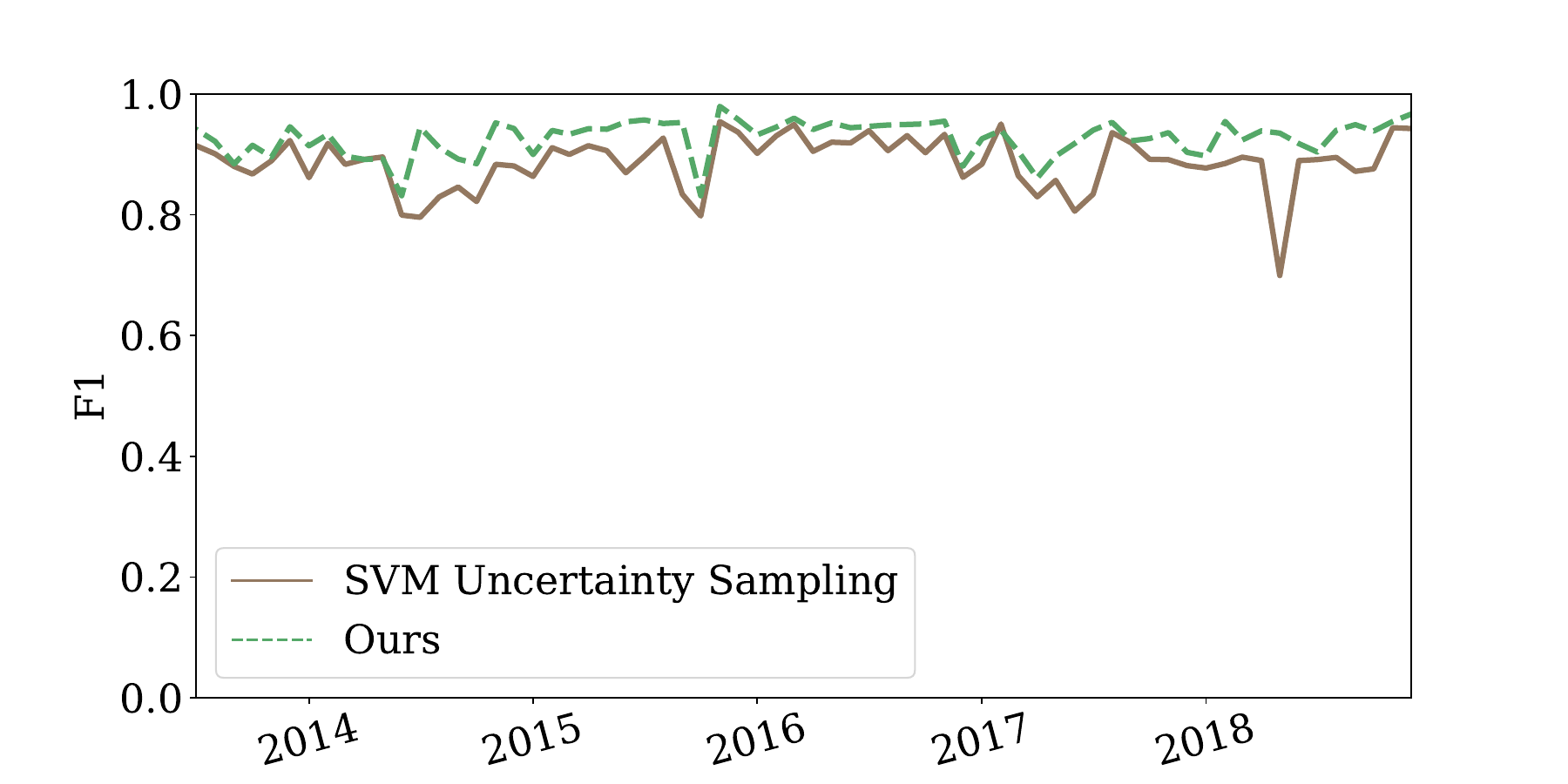}
        \label{fig:ours-svm-unc-cnt200-f1}}
        \caption{Our technique can significantly reduce the FNR and improve the F1 score of the classifier compared to no active learning. Given a fixed budget of 200 samples per month, our technique steadily maintains a lower false negative rate and a higher F1 score than the best baseline active learning method: SVM uncertainty sampling. The best baseline has over 20\% FNR in 9 months during the six years time period, mainly in 2014 and 2015.
        }
        \label{fig:stats_50samples}
\end{figure*}

Table~\ref{tab:main} shows the performance of each classifier, averaged across 2013-07 to 2018-12 on the APIGraph dataset, and across 2020-07 to 2021-12 on the AndroZoo dataset, by false negative rate (FNR), false positive rate (FPR), and F1 score. We observe:
\begin{itemize}[leftmargin=*]
\item On the APIGraph dataset, if we care about achieving an average F1 score of at least 89\%, the best baseline needs 400 samples per month to reach that performance, whereas our technique only needs 50 samples per month. We decrease the labeling cost by 8$\times$.
\item On the APIGraph dataset, when the monthly labeling budget is 50/100/200 samples, our scheme is better in all metrics---FNR, FPR, and F1 scores---compared to the best baseline.
\item On the AndroZoo dataset, given a fixed labeling budget, our method reduces the FNR by $1.6\times$ on average, while maintaining under 1\% FPR.
\item In most cases, the best baseline is linear SVM with uncertainty sampling. It is a simpler classifier than other baselines, which might generalize better when there is concept drift. 
\end{itemize}

In Figure~\ref{fig:stats_50samples}, we visualize the performance of our technique (with 200 samples / month), a baseline with no active learning, and the best baseline with active learning (200 samples / month).
Figure~\ref{fig:ours-plain-cnt200} and Figure~\ref{fig:ours-plain-cnt200-f1} show that our technique significantly improves the false negative rate (FNR) and F1 score of the classifier compared to no active learning. Even the best baseline active learning scheme, SVM with uncertainty sampling, experiences many spikes of high FNR (Figure~\ref{fig:ours-svm-unc-cnt200}) and sudden drops of F1 score (Figure~\ref{fig:ours-svm-unc-cnt200-f1}). In comparison, our technique maintains a more steady performance over six years of data.

\subsection{Comparison against Improved Schemes}
\label{sec:compare-adapted-baselines}

\begin{table*}[ht!]
  \centering
  \small
  \begin{tabular}{c | c | c | c | rrr | rrr }
    \hline
    \multirow{3}{*}{\bf Budget}  & \multirow{3}{*}{\begin{tabular}{@{}c@{}}{\bf Model}\\{\bf Arch}\end{tabular}} &  \multirow{3}{*}{\begin{tabular}{@{}c@{}}{\bf Sample}\\{\bf Selector}\end{tabular}} & \multirow{3}{*}{\begin{tabular}{@{}c@{}}{\bf Warm}\\{\bf or}\\{\bf Cold}\end{tabular}} & \multicolumn{3}{|c}{\begin{tabular}{@{}c@{}}{\bf APIGraph Dataset}\end{tabular}} & \multicolumn{3}{|c}{\begin{tabular}{@{}c@{}}{\bf AndroZoo Dataset}\end{tabular}} \\
     &  &  &  & \multicolumn{3}{|c}{Average Performance (\%)} & \multicolumn{3}{|c}{Average Performance (\%)} \\
     &  &  &  & FNR & FPR & F1 & FNR & FPR & F1 \\
    \hline
    \hline
    \multirow{8}{*}{50} & \multirow{3}{*}{MLP} & Uncertainty & Warm & 21.85 & 0.57 & 84.89 & 48.95 & 0.37 & 62.81 \\
    &  &  CADE OOD & Cold & 17.13 & 0.90 & 86.36 &  43.09 & 0.66 & 67.18 \\
    &  &  CADE OOD & Warm & 13.51 & 1.46 & 86.32 &  43.04 & 0.54 & 67.45 \\
    \cline{2-10}
    &  \multirow{2}{*}{SVM} & \textsc{Transcendent} (cred) & Cold & {\bf 17.48} & {\bf 0.58} & {\bf 87.55} & 49.06 & 0.41 & 62.29 \\
    &   & \textsc{Transcendent} (cred*conf) & Cold & 18.67 & 0.55 & 86.92 & 47.21 & 0.41 & 64.72 \\
    \cline{2-10}
    & {Enc + SVM} & \textsc{Transcendent} (cred) & Cold & 19.75 & 0.59 & 86.02 & {\bf 42.52} & {\bf 0.52} & {\bf 68.56} \\
    \cline{2-10}
     & Ours: & \multirow{2}{*}{Pseudo Loss} & \multirow{2}{*}{Warm} & {\bf 15.15} & {\bf 0.52} & {\bf 89.23}  & {\bf 27.65} & {\bf 0.53} & {\bf 79.92} \\
    & Enc + MLP &  &  & \green{($\downarrow$ 2.33)} & \green{($\downarrow$ 0.06)} & \green{($\uparrow$ 1.68)}  & \green{($\downarrow$ 14.87)} & \red{($\uparrow$ 0.01)} & \green{($\uparrow$ 11.36)} \\
    \hline
    \hline
    \multirow{8}{*}{100} & \multirow{3}{*}{MLP} & Uncertainty & Warm & 17.40 & 0.50 & 87.95 &  47.48 & 0.39 & 64.04\\
    &  &  CADE OOD & Cold & {\bf 14.71} & {\bf 0.79} & {\bf 88.40} &  49.60 & 0.62 & 61.20\\
    &  &  CADE OOD & Warm & 12.35 & 1.41 & 87.22 &  39.33 & 0.48 & 70.92\\
    \cline{2-10}
    &  \multirow{2}{*}{SVM} & \textsc{Transcendent} (cred) & Cold & 17.02 & 0.72 & 87.33 & 43.26 & 0.42 & 67.57  \\
    &   & \textsc{Transcendent} (cred*conf) & Cold & 17.71 & 0.50 & 87.75 &  44.04 & 0.40 & 66.93  \\
    \cline{2-10}
    & {Enc + SVM} & \textsc{Transcendent} (cred) &  Cold & 17.03 & 0.54 & 87.96 & {\bf 34.85} & {\bf 0.52} & {\bf 74.97}  \\
    \cline{2-10}
     & Ours: & \multirow{2}{*}{Pseudo Loss}  & \multirow{2}{*}{Warm}   & {\bf 13.69} & {\bf 0.44} & {\bf 90.42} & {\bf 27.35} & {\bf 0.41} & {\bf 80.07}\\
    & Enc + MLP &  &  & \green{($\downarrow$ 1.02)} &  \green{($\downarrow$ 0.35)} & \green{($\uparrow$ 2.02)} & \green{($\downarrow$ 7.50)} & \green{($\downarrow$ 0.11)} & \green{($\uparrow$ 5.10)}  \\
    \hline
    \hline
    \multirow{8}{*}{200} & \multirow{3}{*}{MLP} & Uncertainty & Warm & 15.87 & 0.59 & 88.53 & 40.52 & 0.49 & 70.04 \\
    &  &  CADE OOD & Cold & 13.25 & 0.77 & 89.26 &  41.99 & 0.68 & 67.70 \\
    &  &  CADE OOD & Warm & 11.78 & 0.80 & 89.99 &  40.16 & 0.46 & 71.15 \\
    \cline{2-10}
    &  \multirow{2}{*}{SVM} & \textsc{Transcendent} (cred) & Cold & 16.15 & 0.61 & 88.22 & 40.85 &  0.38 & 69.89 \\
    &   & \textsc{Transcendent} (cred*conf) & Cold & 18.04 & 0.48 & 87.66 & 38.25 & 0.42 & 71.08 \\
    \cline{2-10}
    & {Enc + SVM} & \textsc{Transcendent} (cred) & Cold & {\bf 13.45} & {\bf 0.52} & {\bf 90.17} &  {\bf 28.54} & {\bf 0.50} & {\bf 80.26} \\
    \cline{2-10}
     & Ours: & \multirow{2}{*}{Pseudo Loss} & \multirow{2}{*}{Warm} & {\bf 9.42} & {\bf 0.48} & {\bf 92.72}  & {\bf 27.67} & {\bf 0.39} & {\bf 80.51}\\
    & Enc + MLP &  &  & \green{($\downarrow$ 4.03)} & \green{($\downarrow$ 0.04)} & \green{($\uparrow$ 2.55)}  & \green{($\downarrow$ 0.87)} & \green{($\downarrow$ 0.11)} & \green{($\uparrow$ 0.25)}\\
    \hline
    \hline
    \multirow{8}{*}{400} & \multirow{3}{*}{MLP} & Uncertainty & Warm & 14.74 & 0.59 & 89.21 & 33.32 & 0.48 & 75.52 \\
    &  &  CADE OOD & Cold & 11.09 & 1.09 & 89.06 &  29.78 & 0.63 & 77.89 \\
    &  &  CADE OOD & Warm & 11.01 & 0.76 & 90.55 &  43.10 & 0.37 & 67.99 \\
    \cline{2-10}
    &  \multirow{2}{*}{SVM} & \textsc{Transcendent} (cred) & Cold & 15.46 & 0.60 & 88.71 & 36.99 & 0.40 & 72.44 \\
    &   & \textsc{Transcendent} (cred*conf) & Cold & 17.45 & 0.50 & 87.90 & 37.11 & 0.38 & 72.52 \\
    \cline{2-10}
    & {Enc + SVM} & \textsc{Transcendent} (cred) & Cold & {\bf 11.30} & {\bf 0.52} & {\bf 91.46} &  {\bf 27.86} & {\bf 0.45} & {\bf 80.84} \\
    \cline{2-10}
     & Ours: & \multirow{2}{*}{Pseudo Loss} & \multirow{2}{*}{Warm} & {\bf 7.84} & {\bf 0.50} & {\bf 93.50}  & {\bf 21.49} & {\bf 0.31} & {\bf 85.81} \\
    & Enc + MLP &  &  & \green{($\downarrow$ 3.46)} & \green{($\downarrow$ 0.02)} & \green{($\uparrow$ 2.04)}  & \green{($\downarrow$ 6.37)} & \green{($\downarrow$ 0.14)} & \green{($\uparrow$ 4.97)} \\
    \hline
  \end{tabular}
  \caption{Given a fixed monthly labeling budget, we compute the average FNR, FPR, and F1 for improved  active learning techniques and our method. On the APIGraph dataset, our method performs better than improved schemes in all metrics. On the AndroZoo dataset, we reduce the FNR by $1.3\times$ on average while maintaining under 1\% FPR.
  }
  \label{tab:adapted-baselines}
\end{table*}

\subsubsection{Improved Active Learning Schemes}
\label{sec:adapted-baseline-setup}

We compare to several active learning schemes that have not been previously proposed or evaluated in the literature, but that are adapted from previously published schemes or with several of our improvement applied.  This allows us to gain insight into the contribution of each of our ideas, and we show evidence that our full scheme does better than any of these alternatives.  In particular, we evaluate two schemes that are based on a previously published method for drift detection (\textsc{Transcendent}), adapted to support  active learning; and we evaluate several schemes that extend previously published work with some of our new ideas, including warm-start uncertainty sampling (where the classifier is updated each month rather than retrained from scratch) and warm-start CADE with neural networks.

We adapt \textsc{Transcendent}~\cite{barbero2022transcending} to active learning.
\textsc{Transcendent}~\cite{barbero2022transcending} was originally designed to support classification with rejection, so that the classifier can decline to make any prediction for samples that appear to have drifted.
In particular, they construct two scores to recognize drifted samples: credibility and confidence.
Given a new test sample, they first compute the non-conformity score of the sample, representing how dissimilar it is from the training set. Given the predicted label of the test sample, they find the set of calibration data points with the same ground truth label. Then, they compute credibility as the percentage of samples in the calibration set that have higher non-conformity scores than the test sample.
% , if they have the same label as the predicted label.
They compute confidence as one minus the credibility of the opposite label. A lower credibility score or a lower confidence score means the test sample is more likely to have drifted.

We design two active learning sample selectors based on \textsc{Transcendent}. The first one uses only the credibility score: samples with the lowest credibility scores are prioritized. The second one uses both credibility and confidence: we multiply the credibility and confidence, and samples with the lowest score are prioritized. To compute non-conformity scores, we use Cross-Conformal Evaluator (CCE) with 10-fold cross validation, with details in Appendix~\ref{appendix:details-transcendent}.

To the best of our knowledge, these two sample selectors have not been documented in published research papers. The most related papers BODMAS~\cite{yang2021bodmas} and CADE~\cite{yang2021cade} experimented with using the non-conformity score to select samples for active learning. They sort samples by credibility first, and then use confidence to break ties.\footnote{This was confirmed via communication with the authors.} This is different from our sample selectors.

We evaluate these \textsc{Transcendent}-derived sample selectors with a binary SVM classifier, trained from the input features. We also apply the \textsc{Transcendent} credibility score sample selector to the embedding space learned by hierarchical contrastive learning (Equation~\ref{eq:lhc}), and train a binary SVM classifier on these embeddings.
We also evaluate improved variants of NN uncertainty sampling and CADE OOD sampling, improved with the engineering insights from Section~\ref{sec:engineering-lessons}, to help us separate out the benefit from engineering improvements vs our hierarchical contrastive classifier and pseudo loss.

We improve CADE to make it more suitable for deep active learning. CADE uses a contrastive autoencoder to learn embeddings and build a similarity measure, but CADE's classifier takes the original features as input, not the embedding produced by the encoder.  Our insight is that it is better for the classifier to use the embedding as input rather than the original features, so we improve CADE in this way. We also replace CADE's SVM classifier with a neural network, which performed better in our experiments. We examine both a cold-start and warm-start version of CADE, as CADE did not experiment with repeated retraining and thus did not examine this tradeoff, but we found that it made a difference for our scheme (see Section~\ref{sec:cold-vs-warm}). Finally, we modified the architecture of the encoder to further improve performance.

We follow the procedure described in Section~\ref{sec:dataset-split} to find the best hyperparameters to train models from improved active learning schemes, with details in Appendix~\ref{appendix:details-adapted-baselines}.
The details of our model is in Appendix~\ref{appendix:details-our-model}.

\subsubsection{Results}

Table~\ref{tab:adapted-baselines} shows the results of comparing our scheme with these improved schemes. Here are some highlight results:
\begin{itemize}[leftmargin=*]
\item On the AndroZoo dataset, compared to the best improved scheme, our method reduces the FNR by  $1.3\times$ on average, and maintains under 1\% FPR.
In other words, even when improving previously published methods as much as we were able, with all the improvements we could find, our scheme still performs significantly better than prior methods.
\item On the APIGraph dataset, our scheme is better in all metrics, including FNR, FPR, and F1 scores, compared to the best improved scheme.
\item If we exclude our method, \textsc{Transcendent} (cred) applied to the embedding space of hierarchical contrastive learning (Enc + SVM) is the best improved scheme. In one out of eight cases, \textsc{Transcendent} (cred) on the hierarchical embedding space has similar performance as ours, i.e., 200 samples / month for the AndroZoo dataset.
% \item If we exclude our method, MLP uncertainty sampling with warm start is the best improved scheme when the sample budget is 50 and 100; CADE OOD sampling with MLP and warm start is the best improved scheme when sample budget is 10.
\item Our improved CADE schemes are better than the original CADE. For MLP, warm start works better than cold start.
% \item Active learning based upon \textsc{Transcendent} performs similarly to uncertainty sampling (using a SVM classifier for both schemes), with \textsc{Transcendent} performing slightly better. 
\end{itemize}

\subsection{Engineering Lessons}
\label{sec:engineering-lessons}

% This section discusses the engineering lessons we learned when applying active learning for security.

\subsubsection{Hyperparameters for Active Learning}

\textbf{Lesson 1:} concept drift requires a separate hyperparameter tuning procedure for the active learning process.

To learn a fixed classifier, we typically choose hyperparameters of a model such that the performance in the validation set is the best, where the validation set and training set are drawn from the same data distribution. This represents the performance when the classifier is evaluated on the same distribution it is trained on.
However, to be robust against concept drift, we need the classifier to perform well on future data that is from a different distribution. Therefore, we need to use temporally-consistent validation to choose hyperparameters that will perform the best for active learning. We include examples of this phenomenom in Appendix~\ref{appendix:details-hyperparameter-example}.

\subsubsection{Cold Start vs Warm Start}
\label{sec:cold-vs-warm}

\textbf{Lesson 2:} warm start is better than cold start when using deep active learning for malware detection.

In active learning, there are two options to train a new model after labeling new samples: cold start or warm start. Cold start re-initializes the model weights and retrains the model from scratch. Warm start continues training from the previous model weights in each active learning iteration.

Previous works have not studied the benefits of warm start vs cold start.
The active learning experiments from previous security papers use cold start~\cite{zhang2020enhancing,yang2021bodmas,yang2021cade}. Deep active learning papers for image applications have used both cold start~\cite{emam2021active,kongneural,lang2021best} and warm start~\cite{yoo2019learning,zhang2020state}, but they did not find much difference between the two strategies.

% \begin{table}[ht!]
%   \centering
%   \small
% 	\begin{tabular}{c | r r r}
% 		\hline
% 		\multirow{2}{*}{\textbf{Setting}} & \multicolumn{3}{|c}{\textbf{Average (\%)}} \\
%   & FNR & FPR & F1 \\
%   \hline
% 	\multirow{3}{*}{\begin{tabular}{@{}c@{}}{\bf Baseline:}\\{Standard Classifier}\\{+ Standard Uncertainty Sampling AL}\end{tabular}}	& \multirow{3}{*}{14.07} & \multirow{3}{*}{0.86} & \multirow{3}{*}{88.47} \\
%             & & & \\	
%             & & & \\
% 		\hline
%         \multirow{2}{*}{\begin{tabular}{@{}c@{}}{Hierarchical Contrastive Learning}\\{+ Standard Uncertainty Sampling AL}\end{tabular}}	& \multirow{2}{*}{11.87} & \multirow{2}{*}{0.45} & \multirow{2}{*}{91.47} \\
%             & & & \\
%             \hline
%         \multirow{2}{*}{\begin{tabular}{@{}c@{}}{Standard Contrastive Learning}\\{+ Pseudo Loss Sampling AL}\end{tabular}}	& \multirow{2}{*}{11.01} & \multirow{2}{*}{0.53} & \multirow{2}{*}{91.56} \\
%             & & & \\
%             \hline
%         \multirow{3}{*}{\begin{tabular}{@{}c@{}}{\bf Ours:}\\{Hierarchical Contrastive Learning}\\{+ Pseudo Loss Sampling AL}\end{tabular}}	& \multirow{3}{*}{9.42} & \multirow{3}{*}{0.48} & \multirow{3}{*}{92.72} \\
%             & & & \\
%             & & & \\
%             \hline
% 	\end{tabular} 
% 	\caption{Backup Ablation Study Table.}
% 	\label{tab:ablation}
% \end{table}

\begin{table*}[ht!]
  \centering
  \small
	\begin{tabular}{c | c | c | r r r}
		\hline
		\multirow{2}{*}{\textbf{Setting}} & \multirow{2}{*}{\textbf{Classifier}} & \multirow{2}{*}{\textbf{Active Learning Sample Selector}} & \multicolumn{3}{|c}{\textbf{Average (\%)}} \\
  &  &  & FNR & FPR & F1 \\
  \hline
	Baseline & Binary SVM & Uncertainty Sampling & 14.07 & 0.86 & 88.47 \\
        \hline
 	% Contrastive Learning Baseline & Standard Contrastive Learning & Uncertainty Sampling &  &  &  \\
  %       \hline
            \multirow{4}{*}{\begin{tabular}{@{}c@{}}{New Hierarchical}\\{Contrastive Learning}\end{tabular}} & \multirow{4}{*}{New Hierarchical Contrastive Classifier} & \multirow{2}{*}{Uncertainty Sampling} & \multirow{2}{*}{11.87} & \multirow{2}{*}{0.45} & \multirow{2}{*}{91.47} \\
            & & & & & \\
            \cline{3-6}
            & & \textsc{Transcendent} (cred) & 12.27 & 0.53 & 90.99 \\
            \cline{3-6}
            & & \textsc{Transcendent} (cred*conf) & 12.78 & 0.47 & 90.85 \\
            \hline
            New Pseudo Loss Sampling &  Contrastive Classifier & New Pseudo Loss Sampling & 11.01 & 0.53 & 91.56 \\
		\hline
            Ours & New Hierarchical Contrastive Classifier & New Pseudo Loss Sampling & 9.42 & 0.48 & 92.72 \\
            \hline
	\end{tabular} 
	\caption{Combining our two main ideas, new hierarchical contrastive classifier and new pseudo loss sampling, is better than either one on its own. When using both techniques together, we achieve 92.72\% F1 score, higher than using either technique on its own.}
	\label{tab:ablation}
\end{table*}

We find that warm start is better than cold start when using deep active learning for Android malware detection. The main reason is sample imbalance: there are very few newly labeled samples, compared to a large amount of initial training samples.
Several past works~\cite{zhang2020enhancing,yang2021bodmas} have trained the first classifier using one year of labeled samples, containing 30K apps, then labelled a few of the new incoming samples every month. If we label 5\% of new samples every month, that the new samples will be less than 1\% of the training set.
During active learning, we add new samples to the training set and continue training from the previous model weights. Therefore, batches from the new training set typically contain a mix of old and new samples.
Since new samples might represent the trend of concept drift, it is beneficial for the classifier to learn more from the newer samples than the older ones, but does not forget about the oldest samples. Warm start can address the sample imbalance issue. When we continue training a new model from previously learned weights, newly labeled samples have the largest loss values and thus largest gradients, previously labeled samples have relatively smaller loss values, and samples from the initial training set have the smallest loss values. Since newly labeled samples have the largest gradients, this enables the model to learn more from recently labeled samples and adapt to the trend of concept drift.

We use warm start to train our hierarchical contrastive classifier during active learning.
We use the time-consistent validation split to find the best hyperparameters for warm start.We also use the warm start idea to improve deep active learning methods and compare against improved baselines in Section~\ref{sec:compare-adapted-baselines}. This include uncertainty sampling for neural networks and using CADE OOD sample selector to continuously train a neural network model.

\subsection{Ablation Study}

We conduct an ablation study to understand the improvements from the two components in our scheme: the new hierarchical contrastive classifier (Section~\ref{sec:hierarchical-contrastive-classifier}) and the new pseudo loss sample selector (Section~\ref{sec:pseudo-loss}).
Accordingly, we compare: 1) a baseline with neither component (binary SVM with uncertainty sampling), 2) just a hierarchical contrastive classifier without our new sample selector, 3) just our new sample selector, without a hierarchical contrastive classifer (instead we use a contrastive classifier, but no hierarchy), 4) our full method, with both components.
We evaluate on the APIGraph dataset, with 200 samples / month budget. We use warm start to train all schemes.

Our results (Table~\ref{tab:ablation}) show that each component offers improvements, and best results are achieved by combining both components.
Using both techniques achieves 92.72\% F1 score, but using only one of the two can achieve 91.47\% or 91.56\% F1 score. This demonstrates that both components are needed for optimal performance.

We also experiment with a combination of our new hierarchical contrastive classifier and \textsc{Transcendent} sample selectors (instead of our new pseudo loss sample selector). These schemes achieve 90.99\% and 90.85\% F1 score, which is better than the baseline and better than \textsc{Transcendent} on the input feature space, but not as good as our full method.

\section{Case Study}

\begin{figure}[t!]
    \centering
    \includegraphics[width=0.47\textwidth]{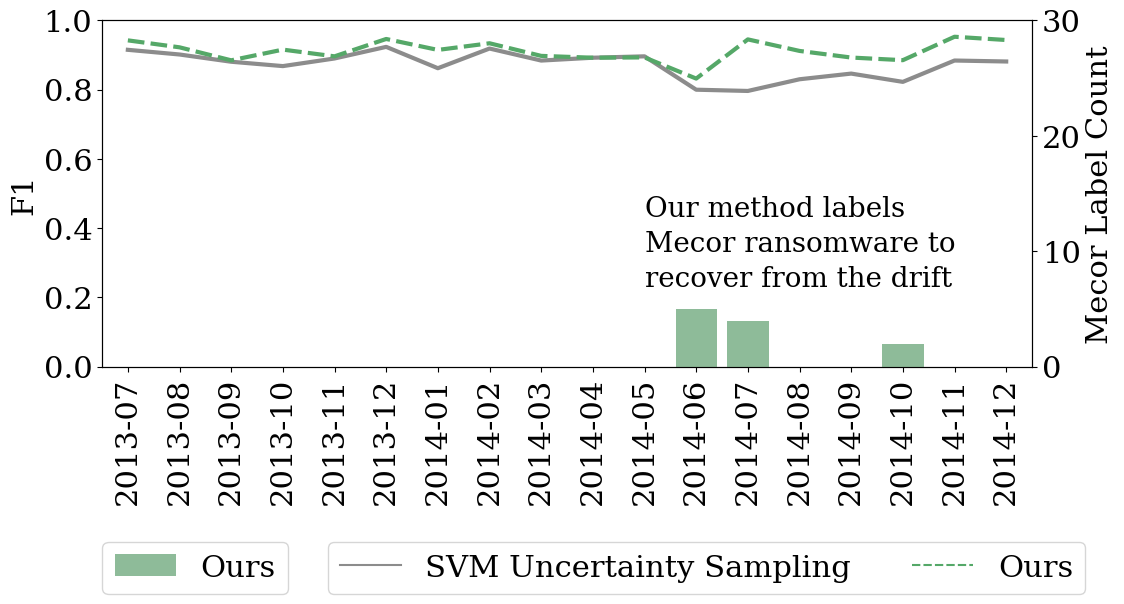}
    \caption{The lines are F1 scores, and the bars are the number of Mecor ransomware samples labeled every month. In 2014-06, Mecor is the family with worst false negative rate. Both SVM and our model have a FNR of 100\% in this family, which causes the F1 score drop to $0.79$ and $0.83$, respectively. Our active learning scheme is able to select Mecor samples and label them as this happens, immediately recovering from the concept drift back to 0.94 F1 score. However, SVM uncertainty sampling fails to select and Mecor samples and continues to have poor model performance in the rest of 2014.}
    \label{fig:ransomware-case}
\end{figure}

In this section, we use a case study to illustrate why our scheme can maintain better performance than the best prior method for active learning.

Even the best baseline method, SVM with uncertainty sampling, cannot avoid many spikes of high false negative rates (FNR) as shown in Figure~\ref{fig:ours-svm-unc-cnt200}.
Figure~\ref{fig:ransomware-case} shows that, in June 2014, for both SVM and our scheme, the F1 score has dropped to $0.79$ and $0.83$ respectively, and the FNR is over 28\%. After looking into the samples, we find that 47\% of the false negative samples are in a ransomware family Mecor, and 100\% of the Mecor samples were misclassified by both SVM and our model. However, our scheme is able to quickly recover from the drift, by selecting Mecor samples in June, July, and October 2014 for labeling. Since we add these samples to the training set and continuously train our classifier, the F1 score of our model immediately goes back to 0.94 in July 2014. In comparison, the SVM uncertainty sampling scheme fails to select any Mecor ransomware samples despite the high FNR for the family, and the F1 score remains low for the rest of 2014.

\section{Discussion}

All machine learning based detection schemes are subject to evasion attacks, for instance using adversarial examples or even simpler methods of evasion (e.g., obfuscation or packing).  It is an open challenge for the field how to solve this problem.  As a machine-learning-based scheme, we inherit these same challenges.  It is beyond the scope of this paper to address this challenge. One potential direction is to extract dynamic features from the apps, or use a combination of static and dynamic features to be more robust against evasion attacks.

% Adversarial machine learning research has demonstrated various ways for an attacker to evade ML-based malware detectors. While all of these are potential risks to our technique since we use deep learning, we would like to discuss the particular evasion risk introduced by the active learning component in our framework.

Continuous learning introduces new risks of poisoning attacks, where an attacker may be able to carefully craft malicious samples and introduce them into the training process.
Clean-label poisoning attacks may be especially dangerous, because they do not require any misbehavior or malice on the part of analysts~\cite{shafahi2018poison}. The attacker can submit carefully crafted Android apps, hope that they have high pseudo loss values so our sample selector will choose them for human labels, and then let analysts generate clean labels. Even if the poisoning apps have the correct label, they may slowly influence the decision boundary of the classifier, and allow other malware apps evade the detection.  All active learning schemes in the literature---including ours---share this potential risk, and it is an open problem how to defend active learning against poisoning attacks.

%Perhaps using repackaged apps is an option for the clean label attack, since repackaged apps add malicious components into the benign app, so they may be similar to benign apps in the original feature space. However, if repackaged apps are correctly labeled as malware by human analysts, contrastive learning might still be able to separate them from benign apps in the embedding space, so our technique has the potential to be robust against the clean label attack. Unfortunately we do not currently have a dataset of repackaged apps to study this, so we will leave this for future work.

% \yizheng{500: 89\% F1; 100: 90\% F1; 200 budget, 92\% F1 score; 400 budget, 93.5\% F1 score.}

We show that with 50 samples per month labeling budget, our technique can achieve 89\% F1 score. In our dataset, 50 samples is 1\% of all apps in a month. To the best of our knowledge, our classifier performance with 1\% labeling budget is the best result compared to the literature of using active learning for Android malware detection. Android malware classification can achieve 99\% F1 score when there is no concept drift. But with concept drift, the performance gap is still quite large, even with our best techniques. It would be great to reach 95\% F1 score with 1\% labeling budget, or to narrow this gap. We suggest it as a valuable open problem for future research to identify new methods that close this gap.  One potential direction might be to study a richer set of features. When there is no concept drift, DREBIN features have been very effective, and using richer features does not appear to offer significant improvements. Perhaps richer features would be more useful for the concept drift problem.

Like most prior work in this space, we use the same set of features in every time window. Studying how to periodically choose new features to combat drift is an interesting direction for future work, but beyond the scope of this work. One recent, concurrent work~\cite{chenoverkill} found that adding new features was not effective at addressing concept drift, so new ideas seem needed.

% remove this to save space
%It is an interesting direction for future work to explore this approach or other approaches to further improve  Android malware classifier performance when there is concept drift.

\section{Conclusion}

Our work points a way towards a framework for continuous learning in security, based on hierarchical contrastive classifiers and active learning with pseudo loss uncertainty scores. We have validated this approach on Android malware classification and shown that it provides improvements over all prior methods.  We speculate that it might be useful for other security tasks as well.

\section*{Acknowledgements}

We thank Limin Yang for discussions of CADE and BODMAS. This work was supported by Google through the ASPIRE program, by an Amazon research award, by the National Science Foundation through award CNS-2154873, by C3.AI's Digital Transformation Institute, and by the Center for AI Safety Compute Cluster. Any opinions, findings, and conclusions
or recommendations expressed in this material are those of the
authors and do not necessarily reflect the views of the sponsors.

\small
\bibliographystyle{abbrv}
\bibliography{drift}

\appendix
\normalsize
\balance
\section{Details about Initial Training Samples}
\label{appendix:details-training-samples}

We start with the following set of initial training samples to train all models before doing active learning. We extract DREBIN features from both datasets. On the APIGraph dataset, we train on 2012 data, containing 3,061 malicious apps and 27,472 benign apps. We select features with larger than 0.001 variance. We end up with 1,159 selected features. On the AndroZoo dataset, we train on the 2019 data, consisting of 4,542 malicious apps and 40,947 benign apps. We increase the variance threshold such that we select under 20K features with the largest variance. We end up with 16,978 features with the largest variance.

\section{Details about Our Model}
\label{appendix:details-our-model}

Our encoder subnetwork has fully connected layers with ReLU activation. The encoder layers gradually reduce the input features to a 128-dimension embedding space, i.e., `512-384-256-128'. The classifier subnetwork uses two hidden layers, each with 100 neurons and ReLU activation, and two output neurons normalized with Softmax. The two outputs represents the normalized prediction scores for benign and malicious classes, respectively.
We train our encoder-classifier model end-to-end using the loss function in Equation~\eqref{eq:training_loss}. We use batch size 1,024, since a larger batch size produces more pairs for contrastive learning, which typically performs better than smaller batch sizes.

The candidate hyperparameters to train our model are the following. We consider two optimizers: SGD and Adam; 4 initial learning rate choices: 0.001, 0.003, 0.005, 0.007; 3 learning rate schedulers: cosine annealing learning rate decay without restart, step-based learning rate decay by a factor of 0.95 or 0.5 every 10 epochs; 4 choices for first classifier epochs: 100, 150, 200, 250; warm start optimizers: SGD and Adam; warm start learning rate: 1\%, 5\% of the initial learning rate, same learning rate decay as the first model; warm start epochs: 50, 100.

% al_adam/035_warm_lr0.005_sgd_step_0.95_e100_wlr5e-05_we50_test_2013-01_2014-12_cnt50.csv
We use the following hyperparameters for the APIGraph dataset: use SGD optimizer to train the first model, initial learning rate 0.003, step-based learning rate decay by a factor of 0.95 every 10 epochs, 250 training epochs; during warm start, use Adam optimizer, $1.5 * 10^{-4}$ warm learning rate (5\% of the initial learning rate), 100 warm training epochs after adding the new samples from every month.

We use the following hyperparameters for the AndroZoo dataset: use SGD optimizer to train the first model, initial learning rate 0.001, step-based learning rate decay by a factor of 0.5 every 10 epochs, 200 training epochs; during warm start, use Adam optimizer, $1 * 10^{-5}$ warm learning rate (1\% of the initial learning rate), 50 warm training epochs after adding the new samples from every month.

Using one NVIDIA A5000 GPU, training or updating a model takes 10 minutes for the APIGraph dataset. Training and/or testing time is generally fast enough that it is unlikely to be a barrier to deployment; accuracy is the primary challenge.

\section{Details about Baselines}
\label{appendix:details-baselines}

For MLP with uncertainty sampling, we use the same architecture as our classification subnetwork: two hidden layers, each with 100 neurons and ReLU activation, and two output neurons normalized with Softmax. We use batch size 32, and Adam optimizers. We search for learining rate from 0.0001 to 0.0009 with a step size 0.0002; training epochs 25, 50, 75, 100. The best hyperparameters are: 0.007 learning rate and 50 epochs for the APIGraph dataset, 0.001 learning rate and 25 epochs for the AndroZoo dataset.

For SVM with uncertainty sampling, we search for C from the set: 0.001, 0.01, 0.1, 1, 10, 100, 1000. The best C is 0.1 for the APIGraph dataset and 0.01 for the AndroZoo dataset. For SVM with CADE OOD sample selection, we use the exact same setup described in the paper, including their model architecture and batch size, and we will adapt and improve their method in Section~\ref{sec:adapted-baseline-setup}. We train the linear SVM classifier with L2 regularization, squared hinge loss, with prediction probabilities calibrated by logistic regression.

For the multiclass MLP, multiclass MLP embedding (+ SVM), we search through learning rate from 0.001 to 0.009 with a step size 0.002, training epochs 25 and 50. The final setting of multiclass MLP for the APIGraph dataset is: 0.001 learning rate and 50 epochs; for the AndroZoo dataset is: 0.003 learning rate and 50 epochs. Since the benign class is the majority, using random batch sampler gives us a degenerate solution of multiclass MLP classifiers that always predict the benign class. Therefore, we randomly select 10 samples from each class within a batch, such that the number of samples are balanced across different classes. We also tried upsampling all classes to have the same number of samples as the benign class, which has the same effect as randomly selecting 10 samples / class.

For SVM used in the multiclass experiments, we search through the same set of C values mentioned above. The best C is 0.1 for the APIGraph dataset; and 0.01 for the AndroZoo dataset.

\begin{table*}[ht!]
  \centering
  \small
	\begin{tabular}{c | c | c c}
		\hline
		\multirow{2}{*}{\textbf{Classifier}} & \multirow{2}{*}{\textbf{Hyperparameters}} & \multicolumn{2}{|c}{\textbf{Average Validation F1 Score (\%)}} \\
  & & 2012 (initial classifier) & 2013-01 to 2013-06 (active learning, uncertainty sampling) \\
  \hline
		\multirow{2}{*}{GBDT}  & trees: 100, max depth: 10 & 99.67\% & 88.62\% \\
              & trees: 60, max depth: 10 & 99.52\% & 89.54\% \\
		\hline
		\multirow{2}{*}{SVM}  & C=1 & 96.27\% & 87.90\% \\
              & C=0.1 & 95.78\% & 89.97\% \\
		\hline
	\end{tabular} 
	\caption{On the APIGraph dataset, the best hyperparameters to train the first classifier may not be the best ones to maintain good performance when there are drifted samples. Using different hyperparameters to train GBDT and SVM, the average validation F1 scores for the initial classifier are very similar. However, the average monthly validation F1 score during six months of active learning in 2013 can be very different. In this example, hyperparameters that generalize better (smaller depth for GBDT, smaller C for SVM) help active learning perform better.}
	\label{tab:params}
\end{table*}

For GBDT with uncertainty sampling, we search for maximal tree depth: 4, 6, 8, 10, 20, 30, 40, 50; boosting rounds: 10, 20, 30, 40, 50, 60, 80, 100. The best choices for APIGraph dataset are max depth 10 and 60 rounds of boosting; and the best ones for AndroZoo dataset are max depth 10 and 80 rounds of boosting.

\section{Details about \textsc{Transcendent} CCE}
\label{appendix:details-transcendent}

We use Cross-Conformal Evaluator (CCE) with 10-fold cross validation for \textsc{Transcendent}, since CCE has the best performance for sample rejection in \textsc{Transcendent}~\cite{barbero2022transcending}.
For each fold of train / validation split, we train a SVM classifier, and compute non-conformity scores for data in the validation set. Then, we can compute the credibility and confidence score of the test sample for that fold.
\textsc{Transcendent}'s implementation of CCE compares the score in each fold to a threshold and takes the majority vote of these comparisons to decide whether to reject the sample~\cite{barbero2022transcending}.
We extend this to a numeric score rather than a binary decision.
We note that \textsc{Transcendent}'s approach is equivalent to computing the median of the scores in each fold, and comparing this median to a threshold.
Therefore, in our active learning scheme, we compute the median credibility and median confidence across the 10 folds for each test sample.

\section{Details about Improved Baselines}
\label{appendix:details-adapted-baselines}

We retrain SVM for two sample selectors: \textsc{Transcendent} (cred), and \textsc{Transcendent} (cred * conf). To retrain SVM, we search for C from the set: 0.001, 0.01, 0.1, 1, 10, 100, 1000. For the APIGraph dataset, the best C for cred is 0.1 , the best C for cred*conf is 0.01. For the AndroZoo dataset, the best C is 0.01.

We adapt MLP uncertainty sampling with warm start.
We search for learning rate from 0.0001 to 0.0009 with a step size 0.0002; training epochs 25, 50, 75, 100; warm learning rate: 1\%, 5\% of the initial learning rate; warm training epochs: 25, 50. The best hyperparameters for the APIGraph dataset are: 0.0009 learning rate, 25 initial training epochs, warm learning rate $4.5*10^{-5}$ (5\% of the initial one), and warm training epochs is 25. The best hyperparameters for the AndroZoo dataset are: 0.0001 learning rate, 25 initial training epochs, warm learning rate $5*10^{-6}$ (5\% of the initial one), and warm training epochs is 25.

We adapt CADE OOD sample selector for MLP in both cold start and warm start. To have a fair comparison, we use the same encoder dimensions as ours, and mirror that as the decoder in CADE. We use the same MLP structure as our classifier subnetwork. We use batch size 1,536. We fix the MLP learning rate (0.001) and training epochs (50), but perform grid search over the same set of parameters for the CADE autoencoder as described in Section~\ref{appendix:details-our-model}. Note that the original active learning experiment in CADE did not tune hyperparameters (Section 6 in ~\cite{yang2021cade}). But we tune hyperparameters including optimizer, initial learning rate, learning rate scheduling, epochs to train the contrastive autoencoder model, warm start learning rate and epochs.

The best cold start parameters for CADE, APIGraph dataset: Adam optimizer, initial learning rate 0.001, step-based decay with a factor 0.95 every 10 epochs, and 150 training epochs. For the AndroZoo dataset: Adam optimizer, initial learning rate 0.001, step-based decay with a factor 0.5 every 10 epochs, and 100 training epochs.

The best warm start parameters for CADE for the APIGraph dataset: Adam optimizer for both initial classifier and active learning; autoencoder: initial learning rate 0.001, cosine annealing learning rate decay without restart, 250 initial training epochs; active learning: for both the autoencoder and MLP, 5\% of initial learning rate for warm start, and 50 warm training epochs.
For the AndroZoo dataset: Adam optimizer for both initial classifier and active learning; autoencoder: initial learning rate 0.001, cosine annealing learning rate decay without restart, 100 initial training epochs; active learning: for both the autoencoder and MLP, 1\% of initial learning rate for warm start, and 50 warm training epochs.

\section{Hyperparameter Examples}
\label{appendix:details-hyperparameter-example}

Table~\ref{tab:params} shows examples where the best hyperparameters to train the first classifier are not the best ones for active learning.

To evaluate the performance of the initial classifier, we randomly separate apps from 2012 data of the APIGraph dataset into five train/validation splits and average the validation F1 score of the classifier over the splits. The third column of Table~\ref{tab:params} shows that different hyperparameters do not make much difference in the validation F1 score for the initial classifier. 

To evaluate the performance of the classifier trained with active learning, we train an initial model on all 2012 data. Then, we use the first six months in 2013 for active learning. We perform uncertainty sampling by adding 50 new samples to the training set every month, retrain the classifier, and evaluate the F1 score with data from the future month. We average the monthly F1 scores to evaluate the performance during active learning. As shown in the last column of Table~\ref{tab:params}, different hyperparameters can make a significant difference to performance from 2013-01 to 2013-06. The best hyperparameters for active learning are not the best to train the initial model, but they are better for generalization. For GBDT, a smaller number of trees makes the model simpler and less prone to overfitting, which makes the model more robust against concept drift. For SVM, a smaller C value allows more classification mistakes when maximizing the margin, which encourages the generalization of the classifier under concept drift.

\end{document}